\documentclass[fleqn,usenatbib]{mnras}

\usepackage[T1]{fontenc}
\usepackage{comment}
\DeclareRobustCommand{\VAN}[3]{#2}
\let\VANthebibliography\thebibliography
\def\thebibliography{\DeclareRobustCommand{\VAN}[3]{##3}\VANthebibliography}
\DeclareUnicodeCharacter{2212}{-} 

\usepackage{graphicx}
\usepackage{amsmath}
\usepackage{amssymb}
\usepackage{float}
\usepackage{hyperref}
\usepackage{comment}
\usepackage{xcolor}
\usepackage{tikz}
\definecolor{lime}{HTML}{A6CE39}
\DeclareRobustCommand{\orcidicon}{\hspace{-3mm}
	\begin{tikzpicture}
		\draw[lime, fill=lime] (0,0) 
		circle [radius=0.16] 
		node[white] {\hspace{0.1mm}{\fontfamily{qag}\selectfont \tiny ID}};
		\draw[white, fill=white] (-0.07,0.1) 
		circle [radius=0.01];
	\end{tikzpicture}
	\hspace{-5mm}
}

\foreach \x in {A, ..., Z}{\expandafter\xdef\csname orcid\x\endcsname{\noexpand\href{https://orcid.org/\csname orcidauthor\x\endcsname}
		{\noexpand\orcidicon}}
}

\title[Quasar Proximity zones]{The need for obscured supermassive black hole growth to explain quasar proximity zones in the epoch of reionization}

\newcommand{\msun}{\mathrm{M}_\odot}
\newcommand{\lya}{Ly~$\alpha$}

\newcommand{\HI}{H\,I}

\newcommand{\HeI}{He\,I}
\newcommand{\HeII}{He\,II}
\newcommand{\HeIII}{He\,III}

\newcommand{\ud}{\mathrm{d}}

\newcommand{\nel}{n_\mathrm{e}}
\newcommand{\nH}{n_\mathrm{H}}
\newcommand{\nHe}{n_\mathrm{He}}
\newcommand{\nHI}{n_\mathrm{HI}}
\newcommand{\nHII}{n_\mathrm{HII}}
\newcommand{\nHeI}{n_\mathrm{HeI}}
\newcommand{\nHeII}{n_\mathrm{HeII}}
\newcommand{\nHeIII}{n_\mathrm{HeIII}}
\newcommand{\xHI}{x_\mathrm{HI}}

\newcommand{\bHI}{\beta_\mathrm{HI}}

\newcommand{\bHeI}{\beta_\mathrm{HeI}}
\newcommand{\bHeII}{\beta_\mathrm{HeII}}

\newcommand{\aHII}{\alpha_\mathrm{HII}}

\newcommand{\aHeII}{\alpha_\mathrm{HeII}}
\newcommand{\aHeIII}{\alpha_\mathrm{HeIII}}

\newcommand{\gHI}{\Gamma_\mathrm{HI}}

\newcommand{\gHeI}{\Gamma_\mathrm{HeI}}
\newcommand{\gHeII}{\Gamma_\mathrm{HeII}}

\newcommand{\gbgHI}{\Gamma^\mathrm{HI}_\mathrm{bg}}

\newcommand{\gbgHeI}{\Gamma^\mathrm{HeI}_\mathrm{bg}}
\newcommand{\gbgHeII}{\Gamma^\mathrm{HeII}_\mathrm{bg}}

\newcommand{\rp}{R_\mathrm{p}}
\newcommand{\tq}{t_\mathrm{q}}

\renewcommand{\theenumi}{\textbf{\Alph{enumi}.}}
\renewcommand{\theenumii}{\textbf{\alph{enumii}.}}

\author[Satyavolu et al.]{{
  Sindhu Satyavolu$^{1} $\thanks{E-mail: sindhu@theory.tifr.res.in}\ \ \orcidA{} \, ,
  Girish Kulkarni$^{1}$\ \ \orcidB{}\, ,
  Laura C.~Keating$^{2,3}$\ \ \orcidC{}\;
  , and Martin G.~Haehnelt$^{4,5}$}\ \ \orcidD{}\, \\ 
  $^{1}$Tata Institute of Fundamental Research, Homi Bhabha Road, Mumbai 400005, India\\
  $^2$Leibniz Institute for Astrophysics Potsdam, An der Sternwarte 16, D-14482 Potsdam, Germany\\
  $^3$Institute for Astronomy, University of Edinburgh, Blackford Hill, Edinburgh, EH9 3HJ, United Kingdom\\
  $^4$Institute of Astronomy, University of Cambridge, Madingley Road, Cambridge CB3 0HA, UK \\
  $^5$Kavli Institute of Cosmology, University of Cambridge, Madingley Road, Cambridge CB3 0HA, UK\\
}

\usepackage{xcolor}
\definecolor{notecolor}{rgb}{0.8,0,0}
\definecolor{notecolor2}{rgb}{0.8,0.0,0.8}

\date{Accepted ---. Received ---; in original form ---}

\pubyear{2021}
\usepackage{newtxtext,newtxmath}

\begin{document}
\label{firstpage}
\pagerange{\pageref{firstpage}--\pageref{lastpage}}
\maketitle

\begin{abstract}
Proximity zones of high-redshift quasars are unique probes of supermassive black hole formation, but simultaneously explaining proximity zone sizes and black hole masses has proved to be challenging.  We study the robustness of some of the assumptions that are usually made to infer quasar lifetimes from proximity zone sizes.  We show that small proximity zones can be readily explained by quasars that vary in brightness with a short duty cycle of $f_\mathrm{duty}\sim 0.1$ and short bright periods of $t_\mathrm{on}\sim 10^4$~yr, even for long lifetimes. We further show that reconciling this with black hole mass estimates requires the black hole to continue to grow and accrete during its obscured phase.  The consequent obscured fractions of $\gtrsim$ 0.7 or higher are consistent with low-redshift measurements and models of black hole accretion. Such short duty cycles and long obscured phases are also consistent with observations of large proximity zones, thus providing a simple, unified model for proximity zones of all sizes. The large dynamic range of our simulation, and its calibration to the Lyman-$\alpha$ forest, allows us to investigate the influence of the large-scale topology of reionization and the quasar’s host halo mass on proximity zones.  We find that incomplete reionization can impede the growth of proximity zones and make them smaller up to 30\%, but the quasar host halo mass only affects proximity zones weakly and indirectly.  Our work suggests that high-redshift proximity zones can be an effective tool to study quasar variability and black hole growth.
\end{abstract}


\section{Introduction}
\label{sec:intro}

More than 200 quasars with redshift $z>6$ are now known \citep{sarah_e_i_bosman_2021_5510200}, with the highest-redshift quasar observed at a redshift of $z=7.642$ \citep{2021ApJ...907L...1W}.  Estimates of the mass of the central supermassive black holes (SMBHs) of these quasars range from  $10^7$ to $10^{10}\,\msun$ \citep{2003ApJ...587L..15W,2007ApJ...669...32K,2007AJ....134.1150J,2011Natur.474..616M,2013ApJ...779...24V, 2014ApJ...790..145D,2015Natur.518..512W,2018Natur.553..473B, 2019ApJ...880...77O}.  The existence of such massive black holes at a time when the Universe was only a few hundred million years old is an outstanding problem. Current understanding of SMBH formation is that if they have to grow to be as massive as $10^9 \;\msun$ or more within a Gyr after the Big Bang, there can be two possible formation pathways: black holes either grow from supermassive seeds with masses $\sim 10^5\,\msun$, such as those formed via direct collapse, and subsequently accrete at sub-Eddington rates with large radiative efficiency ($\epsilon_{r}>0.1$), or they grow from low-mass seeds with masses $\sim 10^2-10^3\,\msun$, such as those resulting from Population~III stars or runaway mergers in dense star clusters, and continuously accrete gas at super-Eddington rates.  Both alternatives face theoretical hurdles: while direct collapse black holes are rare and form only in special overdense environments, continuous super-Eddington gas accretion with low-mass seeds is hard to achieve because of feedback \citep{1984ARA&A..22..471R,2010A&ARv..18..279V,2012Sci...337..544V,2016MNRAS.457.3356V,2016PASA...33....7J,2019ConPh..60..111S,2020ARA&A..58...27I,2022Natur.607...48L,2009MNRAS.396..343R, 2020MNRAS.498.5652K}.

Any constraints on the growth history of high-redshift quasars are therefore valuable ingredients for models of SMBH formation.  An important source of such constraints is the measurement of proximity zone sizes in rest-frame UV spectra of high-redshift quasars.  Proximity zones are narrow regions blueward of the quasar's rest-frame \lya\ wavelength where the intrinsic quasar spectrum is spared from strong, saturated \lya\ absorption by intergalactic hydrogen.  Transmission inside the proximity zones owes its existence to the reduced neutral hydrogen density because of the hydrogen-ionizing radiation emitted by the quasar.  Observationally, the size of the proximity zone has been defined to be the distance from the quasar to the location where the transmitted \lya\ flux, after being smoothed by a 20\;\AA\ box-car filter, drops below 10\% \citep{2006AJ....132..117F,2010ApJ...714..834C,2017ApJ...849...91M,2017ApJ...840...24E,2020ApJ...900...37E,2020ApJ...903...60I}.  
 Then, the sensitivity of the proximity zone size to the quasar lifetime can be understood as follows. In the simplest scenario, ignoring recombinations and assuming the hydrogen density distribution around the quasar to be uniform, the proximity zone size tracks the radius of the ionized region around a quasar, which, before ionization equilibrium is reached, is given by 
\begin{multline}
  R_{\text{ion}} = 21.2~\mathrm{pMpc}\left(\frac{\dot{N}}{10^{57}\mathrm{s}^{-1}}\right)^{1/3} \left(\frac{t_{\text{q}}}{1~\mathrm{Myr}}\right)^{1/3}  \\ \times \left(\frac{n_{\text{H}}}{7\times10^{-5}~\mathrm{cm^{-3}}}\right)^{-1/3}\left(\frac{x_{\text{HI}}}{10^{-4}}\right)^{-1/3},
  \label{eq:rion}
\end{multline}
where $\dot{N}$ is the rate at which hydrogen-ionizing photons are emitted by the quasar, $n_{\ion{H}{}}$ is the hydrogen density, $x_{\text{HI}}$ is the neutral hydrogen fraction, and $t_{\text{q}}$ is the lifetime of the quasar. The observer's definition of the proximity zone size described above is closely related to $R_{\text{ion}}$, although the two sizes are not identical. Equation~(\ref{eq:rion}) suggests that a measurement of $R_\mathrm{ion}$ can be used to constrain the quasar lifetime as well as its hydrogen environment.  In reality, the gas density and the neutral hydrogen fraction are inhomogeneously distributed, and the dependence of $R_{\text{ion}}$ on the photon emission rate and quasar lifetime is complex.  As a result, proximity zone models use cosmological simulations post-processed with radiative transfer codes \citep{2007MNRAS.374..493B,2006IAUJD...7E..17M, 2007MNRAS.376L..34M,2007ApJ...670...39L,2015MNRAS.454..681K,2016ApJ...824..133K, 2020MNRAS.493.1330D,2021ApJ...911...60C}.  Such models can then be used to infer the properties of the inter-galactic medium (IGM) as well as quasar lifetimes.  

Proximity zone sizes have now been measured for a handful of  quasars with $z>6$ \citep{2006AJ....132..117F, 2010ApJ...714..834C,2011Natur.474..616M,2015ApJ...801L..11V,2017ApJ...849...91M, 2017ApJ...840...24E,2018Natur.553..473B,2020ApJ...903...60I,2017MNRAS.468.4702R,2021ApJ...909...80B}.  The resultant values range from 10 pMpc to 0.14 pMpc across redshifts 5--7.  The highest redshift at which a proximity zone size has been measured is at $z = 7.54$ \citep{2018Natur.553..473B} for a quasar of magnitude $\mathrm{M}_{1450}=-26.7$ for which the proximity zone size is 1.3~pMpc, a factor of three to four smaller than typical proximity zones measured at redshift $z\sim 6$. These measured proximity zones have been used to estimate lifetimes of redshift-6 quasars to be around $10^6$~yr on average \citep{2017ApJ...840...24E,2020MNRAS.493.1330D,2020ApJ...900...37E,2021ApJ...921...88M}.  Interestingly, \citet{2017ApJ...840...24E} and \cite{2020ApJ...900...37E} reported the discovery of seven quasars with extremely small proximity zone sizes that appear to imply very short quasar lifetimes of about $10^4$~yr.

Lifetimes as small as $10^4$~yr \citep{2017ApJ...840...24E, 2020ApJ...900...37E, 2020ApJ...903...34A} are challenging for SMBH formation models. The Salpeter time \citep{1964ApJ...140..796S}, or the e-folding time, for a black hole growing exponentially at the Eddington limit with a radiative efficiency of 0.1 is $4.5 \times 10^7$ yr. Therefore, if the  quasar lifetime is only $10^4$~yr, which is $\sim 0.005\;t_{\mathrm{Salpeter}}$, then the black hole hardly grows, as 
\begin{equation}
  M_{\mathrm{BH}} = M_{\mathrm{seed}}\; \mathrm{exp}\left(\frac{t_{\mathrm{q}}}{t_{\mathrm{Salpeter}}}\right).
  \label{eq:smbh_rate}
\end{equation}
This requires the  black hole seed to be heavier than even the most massive direct collapse black hole seeds suggested ($\sim10^6\,\msun$; \citealt{2020ARA&A..58...27I}).

It is therefore important to critically examine the inference of quasar lifetimes from observed proximity zone sizes.  Three important uncertainties that affect this inference are the large-scale ionization environment of the quasars, their large-scale cosmological density environment, and quasar variability.

First, the redshift range inhabited by these quasars is also witness to a rapid, large-scale change in the ionization state of the Universe due to reionization.  While the details of how reionization occurs and what causes it remain uncertain, it has been argued recently that the spatial fluctuations in the $z\sim 5$--$6$ \lya\ forest require reionization to end as late as $z\sim 5.3$ \citep{2019MNRAS.485L..24K}. It has also been a common, conservative, assumption in the literature that reionization is caused by the hydrogen-ionizing radiation produced by young massive stars in star-forming galaxies.  As a result, the ionization and thermal state of the medium in which a proximity zone is produced is already affected by a complex interplay of stellar radiation and the intergalactic hydrogen. Therefore, it is necessary to include a proper model of reionization while simulating proximity zones. Previous models of quasar proximity zones often made simplifying assumptions about the ionization and thermal environment of high-redshift quasars.  They either assumed the initial ionization state around the quasar to be set by a homogeneous UV background or to be uniformly ionized or neutral \citep{2007MNRAS.374..493B, 2007MNRAS.376L..34M, 2011MNRAS.416L..70B, 2015MNRAS.454..681K, 2017ApJ...840...24E}.  \citet{2007ApJ...670...39L} were the first to point out that this assumption would not be representative of the inhomogeneous IGM at $z \sim 6$. They performed three-dimensional radiative transfer simulations to obtain the patchy UV background at this redshift, but found that the patchy ionization structure of the IGM around the quasar has little effect on quasar proximity zone sizes as quasars tend to reside in regions that are already ionized.  \citet{2018ApJ...864..142D} have modelled proximity zones and damping wings in two $z>7$ quasars with the help of semi-numerical reionization simulations. Recently, \citet{2021ApJ...911...60C} also implemented patchy ionization in their quasar proximity zone models by means of the CROC radiative transfer simulations, although the models considered by them reionize too early to be consistent with \lya\ forest measurements.

\begin{figure*}
  \includegraphics[scale=0.45]{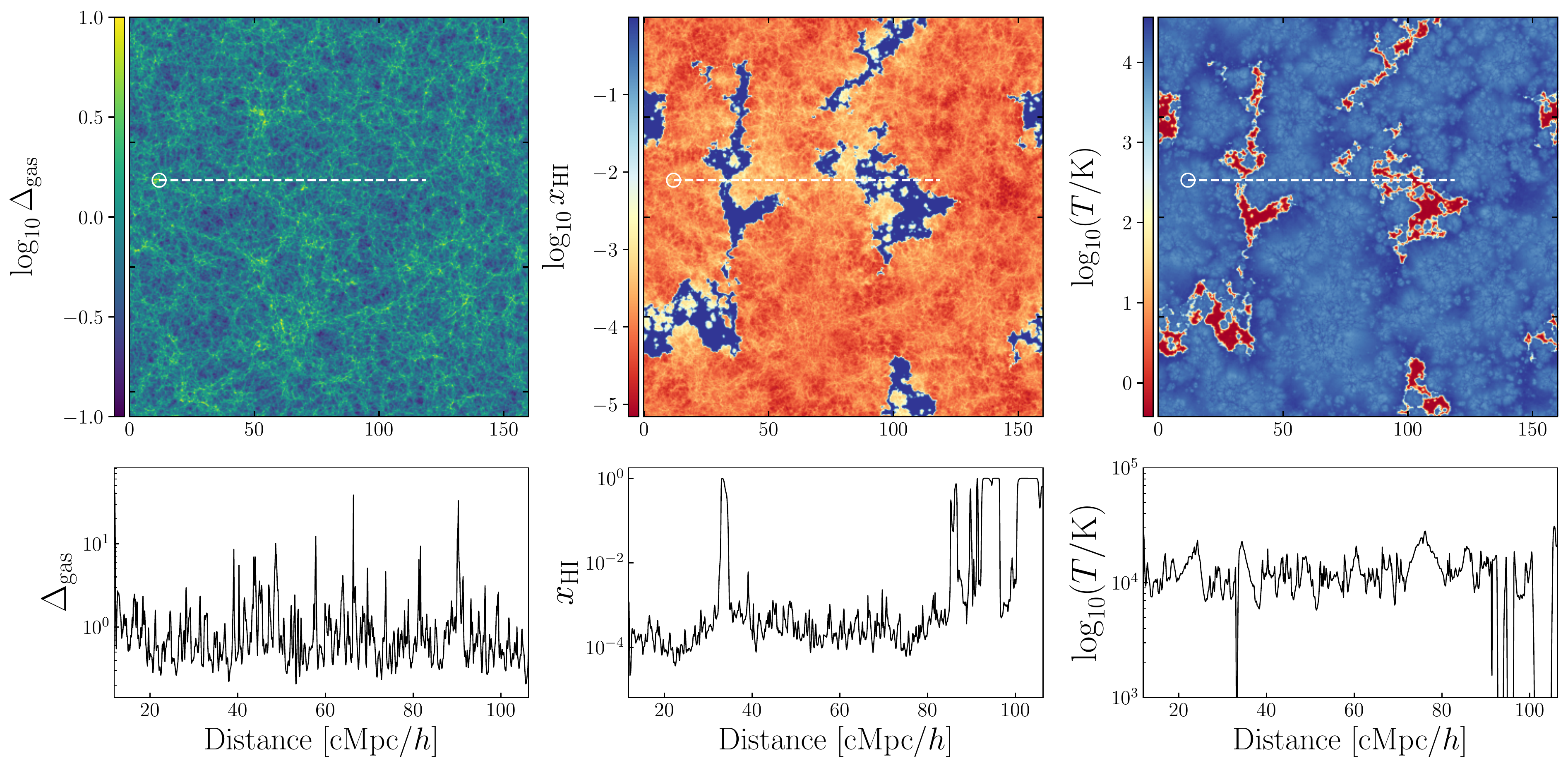}
  \caption{Distribution of the gas density $\Delta_{\text{gas}}$, neutral hydrogen fraction $x_{\text{HI}}$, and gas temperature $T$ at $z=5.95$ from our simulation. The small white circle in the top panels marks the location of a halo of mass $6.97\times10^{11}\;\msun$ .  The bottom panels show the same quantities as the top panel along a one-dimensional skewer drawn from the three-dimensional snapshot, starting from the halo location highlighted in the top panels.  The skewer is also shown in the top panels with a dashed white line. }
  \label{fig:3dsnapshots}
\end{figure*}

Second, similar to the ionization structure of the quasar environment, the uncertain cosmological density structure around high-redshift quasars can also potentially play a role in setting the proximity zone size.  The distribution of host halo masses of high-redshift quasars is not well understood. It is often assumed that the most luminous quasars reside in the most massive halos (e.g., \citealt{2005Natur.435..629S}), but observationally the evidence is uncertain \citep{2007ApJ...654..115C, 2009ApJ...695..809K}. \citet{2007AJ....133.2222S} measured clustering around $z\sim 3$ quasars, which suggested that they lived in massive halos with a minimum mass of $\sim10^{12}\;\msun$.  \citet{2022ApJ...927...65G} have also reported strong clustering of galaxies around redshift $z\sim 4$ quasars.  However, based on the spatial correlation of quasars with protoclusters, \citet{2018PASJ...70S..32U} inferred that luminous quasars around $z\sim4$ do not reside in the most overdense regions.  \citet{2020A&A...642L...1M} found an overdensity of galaxies around a quasar at redshift as high as $z\sim 6.31$. There is no consensus on the overdensity around quasars in simulations either.  \citet{2014MNRAS.439.2146C} report that quasars must reside in the most massive halos in highly over dense regions to be able to grow as massive as $10^{9}M_{\odot}$ by redshift 6 without requiring super-Eddington accretion. The BlueTides simulations \citep{2017MNRAS.467.4243D,2018MNRAS.474..597T} find that massive black holes are formed not in massive halos, but in halos with low tidal fields. They suggest that the most massive black holes should also have formed in environments similar to low mass black holes. \citet{2019MNRAS.489.1206H} use the Horizon-AGN simulation to study the environment of high-redshift quasars and conclude that statistically most massive black holes reside in regions with high galaxy counts. \cite{2013MNRAS.436..315F} used semi-analytic models to study the dark matter environment of quasars and conclude that they live in average mass halos.  \citet{2021ApJ...917...89R} used semi-analytical modelling of the relationship between the quasar luminosity and the host halo mass to predict clustering around high-redshift quasars.  \citet{2015MNRAS.454..681K} were the first to study the role of halo environment on quasar proximity zones.  They argued that proximity zone properties do not depend strongly on the host halo mass of the quasars.

\begin{figure*}
  \includegraphics[scale=0.75]{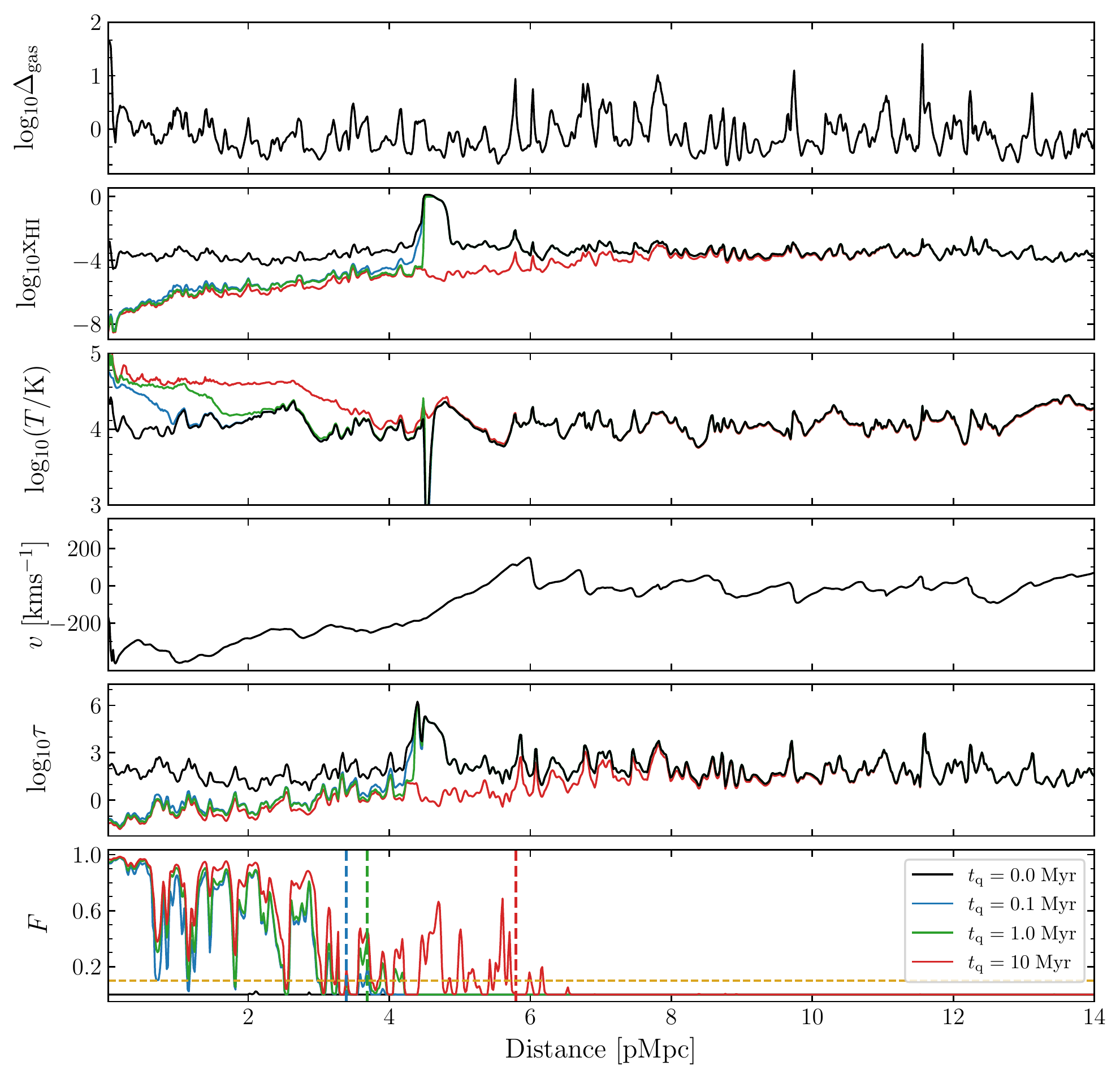} 
  \caption{The result of applying our one-dimensional radiative transfer scheme to the sightline at $z = 5.95$ shown in Figure~\ref{fig:3dsnapshots}.  From top to bottom, the panels show the gas density $\Delta_\mathrm{gas}$, neutral hydrogen fraction $x_\mathrm{HI}$, gas temperature $T$, the peculiar velocity $v$, the \lya\ optical depth $\tau$, and the transmitted flux $F$ along the sightline. Black curves show the quantities before the quasar turns on; colored curves correspond to quasar lifetimes of $t_\mathrm{q}=0.1$, 1, and 10~Myr.  In the bottom panel, the dashed horizontal line shows the 10\% transmission cut-off used to define the proximity zone size, following \citet{2006AJ....132..117F}.  The corresponding proximity zone sizes are demarcated by the vertical dashed lines in this panel.
    \label{fig:examplesightline}}
\end{figure*}

Third, accretion onto SMBHs can be episodic. This variability is conventionally quantified using duty cycles and episodic times, where the duty cycle is the fraction of the quasar lifetime for which a quasar is shining, while the episodic time is the duration of each luminous episode. \citet{2021MNRAS.505.5084W} inferred short episodic times $\lesssim 1$ Myr for four of the thirteen quasars they studied at redshifts $z\sim 3$ from $\ion{He}{ii}$ proximity zones.  These small episodic times were independent of the quasar magnitude, black hole mass, and Eddington ratio, which suggested that their observations must have sampled quasars with short episodic times and large duty cycles. They also remark that if high-redshift quasars follow a similar trend, then most of the black hole growth must have happened during the obscured phase.  \citet{2015MNRAS.451.2517S} estimate that each accreting phase of SMBHs should last around $10^5$ yr based on the time lag between AGN switching on, becoming visible in X-rays, and becoming visible through photoionized narrow lines of the host galaxy. SMBH simulations also suggest that the accretion occurs in episodes shorter than 1 Myr \citep{2017MNRAS.472L.109A,2019MNRAS.483.3488B,2022arXiv220108766M}. \citet{2021ApJ...921...70S} constrained the episodic phases to last for $10^3$--$10^5$~yr based on statistics of `turned-off quasars' and massive galaxies with orphan broad Mg~II emission. So far, the modelling of proximity zones using cosmological simulations post-processed with radiative transfer codes have mostly assumed simple light curves for the quasar, namely the `lightbulb' model where the quasar is shining at a constant luminosity throughout its lifetime. \citet{2020MNRAS.493.1330D} describe an analytical model to predict proximity zone sizes of quasars with blinking light curves as well as for more general light curves and found their model to be in good agreement with their simulations. They conclude that the distribution of proximity zone sizes in such scenarios should allow one to put constraints on the episodic lifetime and duty cycle of the quasar, with their model disfavoring large variations in quasar luminosity below $<10^4$ yr.  

In this work, we investigate the effect of the cosmological density environment of quasars, their large-scale ionization and thermal environment, and episodic accretion activity on quasar proximity zones. We develop and use a one-dimensional radiative transfer scheme together with a high-dynamic-range cosmological radiation transfer simulation of reionization that is calibrated to the \lya\ forest at $z>4$.  The dynamic range of the simulation allows us to span a wide range of host halo masses.  The calibration to the \lya\ forest measurements brings a level of realism to our reionization model.  Next, we confront our model with measurements of proximity zone sizes as well as black hole masses to infer requirements on black hole growth. Our cosmological simulation set up is described in Section~\ref{sec:simulations}.  Section~\ref{sec:methods} presents our radiative transfer method.  Section~\ref{sec:results} contains our main results.  We end with a discussion and a summary of our conclusions in Section~\ref{sec:conclusions}.  Our $\Lambda$CDM cosmological model has $\Omega_\mathrm{b}=0.0482$, $\Omega_\mathrm{m}=0.308$, $\Omega_\Lambda=0.692$, $h=0.678$, $n_\mathrm{s}=0.961$, $\sigma_8=0.829$, and $Y_\mathrm{He}=0.24$ \citep{2014A&A...571A..16P}.

\section{Cosmological Simulations}
\label{sec:simulations}

We use the cosmological simulation previously presented by \citet{2019MNRAS.485L..24K} to set up initial conditions around quasars for redshifts $5<z<9$.  This model consists of a cosmological hydrodynamical simulation developed using P-GADGET-3 (which is a modified version of GADGET-2, described by \citealt{2005MNRAS.364.1105S}), post-processed for three-dimensional radiative transfer using the ATON code \citep{2008MNRAS.387..295A, 2010ApJ...724..244A}.  The box size is 160~cMpc$/h$  with $2048^3$ gas and dark matter particles.  The output of the radiative transfer computation is obtained on a $2048^3$ uniform Cartesian grid with the same box size.  The simulation is run from $z = 99$ to $4$ with initial conditions chosen to be identical to the 160--2048 simulation from the Sherwood Simulation Suite \citep{2017MNRAS.464..897B}.  Snapshots are saved in intervals of 40 Myr.  Sources of ionizing radiation are placed at the centers of masses of friends-of-friends groups with mass $>10^9$~M$_\odot/h$, with the luminosity of each source proportional to the host halo mass with a mass-independent constant of proportionality \citep{2021arXiv210903840H, 2015MNRAS.453.2943C}. As discussed by \citet{2019MNRAS.485L..24K}, this model agrees with the measurement of the distribution of \lya\ opacities at $z > 5$, and also agrees with several other observations such as the CMB optical depth  \citep{2020A&A...641A...6P}, the large-scale radial distribution of galaxies around opaque \lya\ troughs \citep{2020MNRAS.491.1736K,2015MNRAS.447.3402B}, quasar damping wings \citep{2017MNRAS.466.4239G,2019MNRAS.484.5094G,2018ApJ...864..142D,2020ApJ...896...23W}, measurements of the IGM temperature \citep{2020MNRAS.497..906K}, and the luminosity function and clustering of \lya-emitters \citep{2018MNRAS.479.2564W,2019MNRAS.485.1350W}.  Hydrogen reionization ends at $z=5.3$ in this model, with the process half-complete at $z=7$.  This picture continues to be consistent with newer \lya~ opacity measurements \citep{2021arXiv210803699B}.  The locality of the moment-based M1 radiative transfer scheme used in this set-up allows the use of GPUs, which speeds up the radiative transfer computation by a factor of more than 100 relative to CPUs \citep{2010ApJ...724..244A}.  This enables us to enhance the dynamic range of the simulation to include high-mass halos without unduly sacrificing small-scale resolution.  At $z=5.95$, the smallest halo mass resolved in the simulation is $2.32\times10^{8}\;\msun$, while the largest halo mass is $4.59\times10^{12}\;\msun$ .  We refer the reader to \citet{2019MNRAS.485L..24K} and \citet{2020MNRAS.491.1736K} for further details.

For investigating quasar proximity zones, we use simulation snapshots at $z = 5.95$, $6.60$, $7.14$, and $8.15$ in this work.  The volume averaged neutral hydrogen fraction at these redshifts is 0.13, 0.37,0.53 and 0.75 respectively.  As an example, Figure~\ref{fig:3dsnapshots} shows distributions of the gas density $\Delta_{\text{gas}}=\rho_{\text{gas}}/\bar{\rho}_{\text{gas}}$, neutral hydrogen fraction $x_{\text{HI}}$, and gas temperature $T$ at $z=5.95$. The small white circle in the top three panels of Figure~\ref{fig:3dsnapshots} marks the location of a halo of mass $6.97\times10^{11}\;\msun$.  The bottom panels of the figure show the same quantities as the top panel along a one-dimensional skewer drawn from the three-dimensional snapshot, starting from the halo location highlighted in the top panels.  The skewer is also shown in the top panels with a dashed white line, to illustrate the large-scale cosmological environment of this sightline.  Large neutral hydrogen patches of up to 100~cMpc$/h$ in size can be clearly seen at this redshift.  The gas temperature in these regions is less than $10$~K.  Although the neutral hydrogen patches in Figure~\ref{fig:3dsnapshots} are in the deepest voids, making these regions opaque to \lya\ photons, the relationship between \lya\ opacity and large-scale overdensity is non-linear.  This is because regions that are ionized in the most recent past are also in voids, but these regions have higher-than-average temperature, which makes them more \lya\ transparent than overdense regions \citep{2020MNRAS.491.1736K}.  Our simulation set-up allows us to study how this affects quasar proximity zones.  In order to obtain model quasar spectra, we place quasars inside massive halos and perform one-dimensional radiative transfer along skewers starting from the halo, similar to the skewer shown in Figure~\ref{fig:3dsnapshots}.  The advantage of post-processing using one-dimensional radiative transfer as opposed to three-dimensional radiative transfer is that the computational expense is smaller by several orders of magnitude.  Comparisons between three-dimensional and one-dimensional radiative transfer for studying large-scale reionization show little difference in the neutral hydrogen fraction between the two methods \citep{2018MNRAS.476.1741G}.  The details of our one-dimensional radiative transfer method are discussed in the next section.

\section{Quasar spectra}
\label{sec:methods}

We post-process sightlines obtained from the simulation described above with a one-dimensional radiative transfer computation.  The basic equations for the hydrogen and helium ionization chemistry in the presence of photoionization, collisional ionization, and radiative recombination are \citep{2007MNRAS.374..493B, 2013MNRAS.436.2188R}
\begin{align}
  & \frac{\ud\nHII}{\ud t} = \nHI\left(\gHI+\gbgHI+\nel\bHI(T)\right) - \nel\nHII\aHII(T), \label{eq:HI} \\ 
  & \frac{\ud\nHeII}{\ud t} = \nHeI\left(\gHeI+\gbgHeI+\nel\bHeI(T)\right) - \nel\nHeIII\aHeIII(T) \nonumber \\
  & \qquad - \nHeII\left(\gHeII+\gbgHeII+\nel\bHeII(T)\right) - \nel\nHeII\aHeII(T), \label{eq:HeI} \\
  & \frac{\ud\nHeIII}{\ud t} = \nHeII\left(\gHeII+\gbgHeII+\nel\bHeII(T)\right) \nonumber \\ 
  & \qquad - \nel\nHeIII\aHeIII(T). \label{eq:HeIII}
\end{align}
Here, $\Gamma$ and $\Gamma_\mathrm{bg}$ denote the photoionization rates of various species induced by quasars and by background sources, respectively.  The temperature-dependent collisional ionization rates are denoted by $\beta$. Each $\alpha$ refers to the temperature-dependent recombination rates of respective species, and each $n$ denotes their physical number densities from which the electron number density can be computed as $\nel = \nHII +\nHeII + 2\nHeIII$.
Hydrogen and helium abundances were assumed to be of primordial ratio,
\begin{equation}
  \nHe = \frac{Y}{4(1-Y)}\nH,
\end{equation}
where the helium mass fraction is $Y=0.24$.

\begin{figure*}
  \includegraphics[scale=0.45]{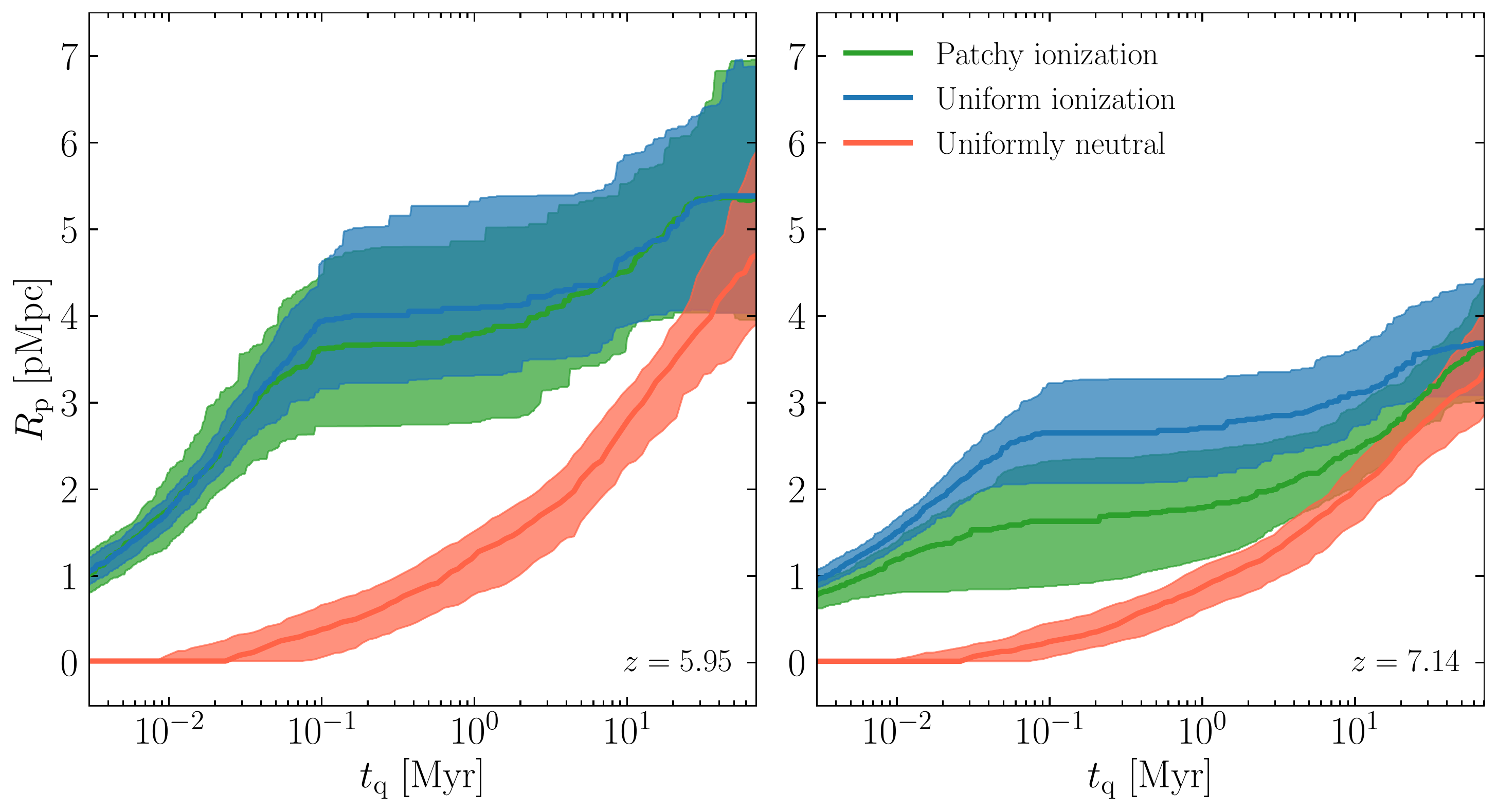}
  \caption{Proximity zone size $\rp$ as a function of quasar lifetime at redshifts $z=5.95$ (left panel) and $z=7.14$ (right panel).  The green curve shows the median evolution of proximity zone size $\rp$ with quasar lifetime $\tq$ from a sample of 100 sightlines with the initial ionization conditions taken from our reionization model.  Shaded region shows the $1 \sigma$ (68.26\% equal-tailed credible interval) scatter across sightlines.  The blue curve and shaded region show the same quantities from a case in which the IGM around the quasar is uniformly ionized.  Similarly, the red curve and shaded regions show results from an initial fully neutral IGM.  \label{fig:ini}}
\end{figure*}

\begin{figure*}
  \includegraphics[scale=0.45]{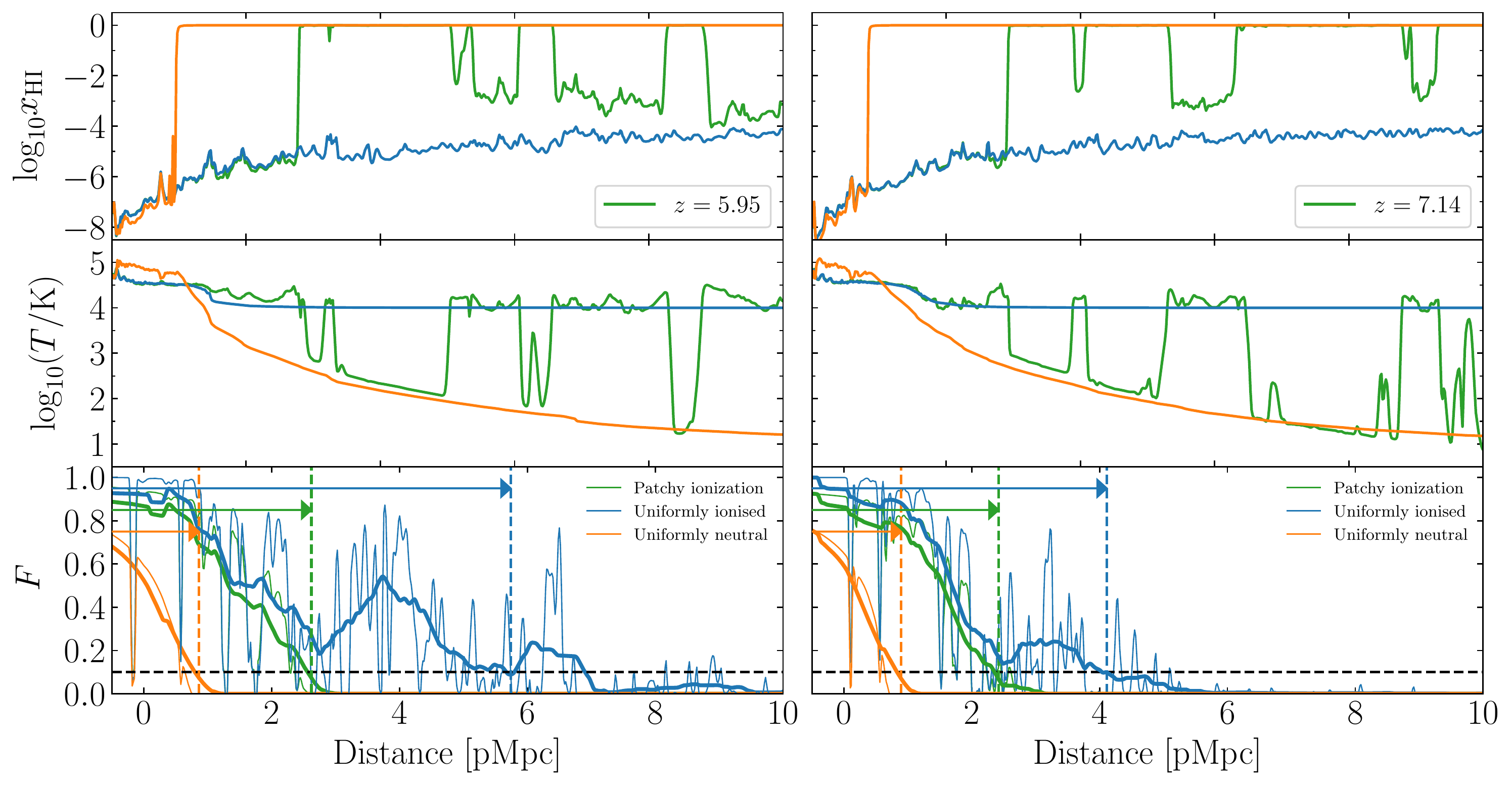}
  \caption{Proximity zone size $\rp$ for different initial conditions: The two panels show neutral hydrogen fraction, temperature and transmitted \lya\ flux at a quasar age $\tq=1$~Myr at two redshifts $z=5.95$ and $7.14$ for the lightbulb model along example sightlines. Green, blue, and orange curves show various quantities from the cases with a patchy initial ionization, uniform initial ionization, and no initial ionization. In the bottom panels, the thick curves show the transmitted flux after being smoothed by a 20~\AA~ boxcar filter and the dashed horizontal line shows the 10\% transmission cut-off used to define the proximity zone size. The corresponding proximity zone sizes are demarcated by the vertical dashed lines in these panels. Neutral hydrogen patches in the IGM impede the growth of the proximity zone relative to the uniformly ionized case.}
  \label{fig:uniform_v_aton_compare_panels}
\end{figure*}

The photoionization rates $\Gamma_{i}$ ($i=\;$\HI, \HeI, \HeII) in a shell of volume $\ud V$ at distance $r$ from the central source are calculated as \citep{2007MNRAS.374..493B}
\begin{equation}
  \Gamma_{i}(r) = \frac{1}{n_i(r)\,\ud V(r)}\int_{\nu_i}^{\infty}\frac{L_\nu}{h_\mathrm{P}\nu}\exp{(-\tau_\nu(r))}\,P_i(r)\,\ud\nu,
  \label{eq:Gamma}
\end{equation}
where the $\nu_i$'s denote the frequencies corresponding to respective ionization thresholds.  The total optical depth is given by $\tau_\nu = \tau^\mathrm{HI}_\nu+ \tau^\mathrm{HeI}_\nu+\tau^\mathrm{HeII}_\nu$, where $\tau^\mathrm{HI}_\nu,\tau^\mathrm{HeI}_\nu $ and $\tau^\mathrm{HeII}_\nu$ are the cumulative sums of the respective optical depths of the three species in all previous shells within radius $r$. These are calculated by summing over the opacities of all shells as 
\begin{equation}
\tau^i_\nu=\sum_{<\,r}\Delta\tau_\nu^i= \sum_{<\,r}  n_i\sigma^i_\nu\Delta r,
\end{equation} 
 where $i=\mathrm{HI},\mathrm{HeI},\mathrm{HeII}$, with $\sigma^i_\nu$  the respective ionization cross-sections, and the sum is over all shells with equal width $\Delta r$. The quantity $P_i$ in Equation~(\ref{eq:Gamma}) represents the conditional probability of a photon being absorbed by species $i$  in the shell at $r$ under the condition that the photon is not absorbed by the other two species in that shell. For the three species, this probability is given by 
\begin{align}
  & P_\mathrm{HI} = p_\mathrm{HI}\,q_\mathrm{HeI}\,q_\mathrm{HeII}\,\left(1-e^{-\Delta\tau_\nu^\mathrm{tot}}\right)/D,\\
  & P_\mathrm{HeI} = q_\mathrm{HI}\,p_\mathrm{HeI}\,q_\mathrm{HeII}\,\left(1-e^{-\Delta\tau_\nu^\mathrm{tot}}\right)/D,\quad \mathrm{and}\\
  & P_\mathrm{HeII} = q_\mathrm{HI}\,q_\mathrm{HeI}\,p_\mathrm{HeII}\,\left(1-e^{-\Delta\tau_\nu^\mathrm{tot}}\right)/D,
\end{align}
where $p_i = 1-e^{-\Delta\tau_{\nu}^{i}}$ is the probability that a photon is absorbed by species $i$ in this shell, and $q_{i}=e^{-\Delta\tau_{\nu}^{i}}$ is the probability that the photon is not absorbed by species $i$ in this shell.  The quantity $1-e^{-\Delta\tau_\nu^\mathrm{tot}}$, with $\Delta\tau^\mathrm{tot}_{\nu}=\sum_{i}\Delta \tau^{i}_{\nu}$ denotes the total probability that a photon is absorbed in the current cell due to all species. The factor $D = p_{\text{HI}}q_{\text{HeI}}q_{\text{HeII}}+q_{\text{HI}}p_{\text{HeI}}q_{\text{HeII}}+q_{\text{HI}}q_{\text{HeI}}p_{\text{HeII}}$ normalizes the probabilities such that the total number of photons absorbed per unit time in a given cell due to all species is \citep{2006NewA...11..374M}
\begin{align}
  \sum_{i} n_{i}(r)\ud V(r)\Gamma_{i} & = \int_{\nu_i}^{\infty} \ud\nu\; \Big[ \frac{L_\nu}{h_\mathrm{P}\nu} \times \exp{(-\tau_\nu(r))}\nonumber \\
   & \times \left(1-e^{-\Delta\tau^{\mathrm{tot}}_\nu(r)}\right)\Big].
\end{align}		
In Equation~(\ref{eq:Gamma}), $L_{\nu}$ is the specific luminosity of the quasar.  This is related to the total number of photons emitted per unit time as 
\begin{equation}
  \dot{N} = \int_{\nu_{\text{HI}}}^{\infty}\frac{L_{\nu}}{h_\mathrm{P}\nu}\ud\nu.
\end{equation}
For a quasar source, we assume the specific luminosity to be a broken power-law in frequency
\begin{equation}
  L_{\nu} = L_{\text{HI}}\left(\frac{\nu}{\nu_{\text{HI}}}\right)^{-\alpha_{s}};\quad \nu>\nu_{\text{HI}}.
\end{equation}
The spectral index $\alpha_s$ is chosen as 1.7 based on the profile of quasars observed around $z\sim 3$ \citep{2015MNRAS.449.4204L}. By assuming a power law with slope $-0.61$, the specific luminosity at hydrogen ionizing edge $L_{\text{HI}}$ can be computed from the specific UV luminosity $L_{1450}$ at 1450\;\AA ~\citep{2015MNRAS.449.4204L}, which in turn can be derived from the observed UV magnitude $M_{1450}$ as 
\begin{equation}
  L_{1450} = 10^{(51.60-M_{1450})/2.5} \mathrm{erg\, s^{-1} Hz^{-1}}.
\end{equation}
We set the photoionization rate due to background sources, $\Gamma_\mathrm{bg}^{i}$, by using the gas density values of our simulation and assuming equilibrium with the IGM before the quasar is turned on \citep{2021ApJ...911...60C}.  This background ionization rate is of the order of $\sim 10^{-12}\,\mathrm{s}^{-1}$. The size of the proximity zone turns out not to have a strong dependence on the background photoionization for the quasar luminosities that we consider \citep[cf.][]{2017ApJ...840...24E, 2020MNRAS.493.1330D}.

\begin{figure}
  \includegraphics[width=\columnwidth]{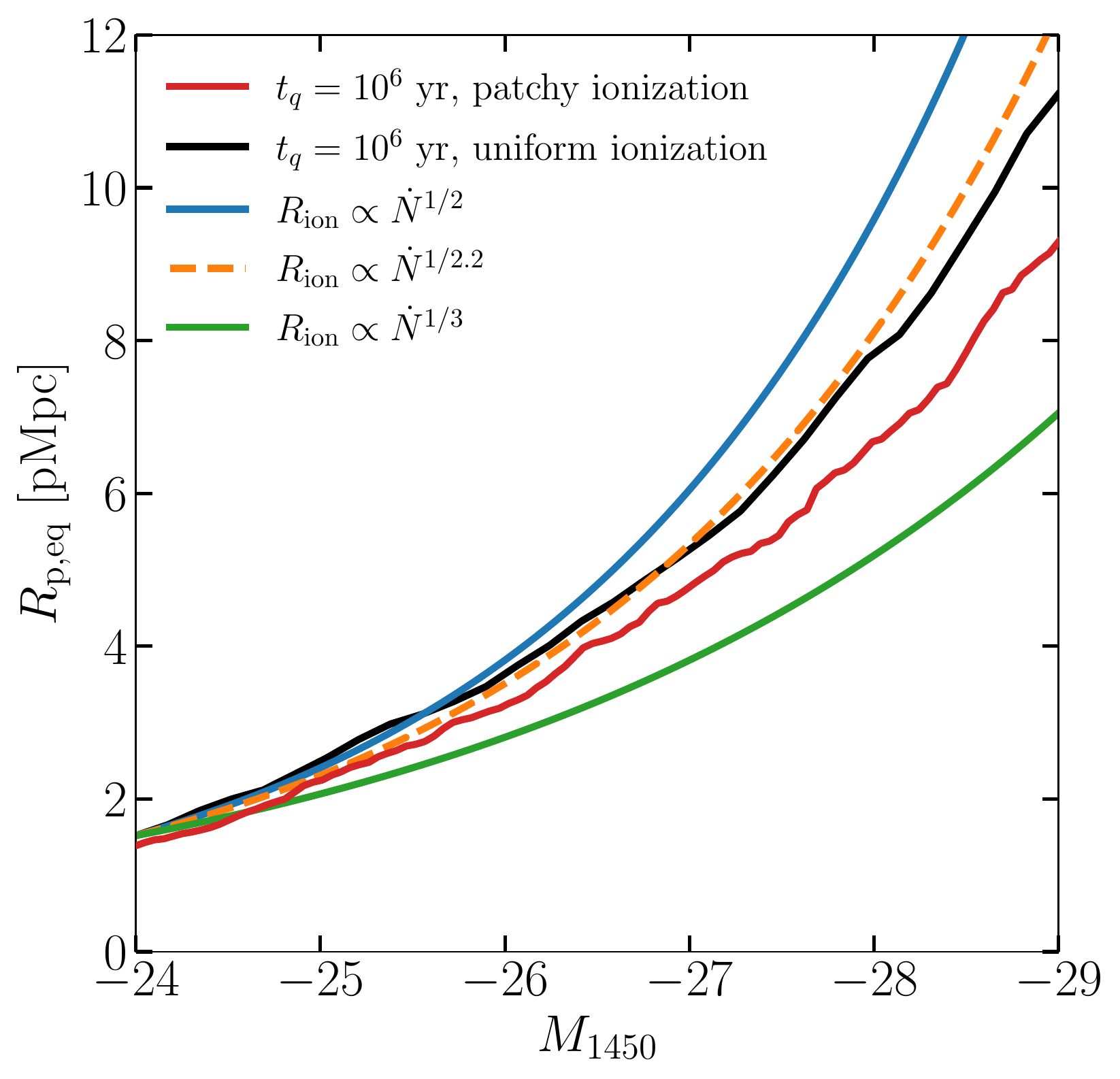}
  \caption{The scaling of the proximity zone size at equilibrium, $R_\mathrm{p,eq}$, with the quasar magnitude at redshift $z=5.95$. Black curve shows the mean $R_\mathrm{p,eq}$ in the case of an initially uniformly ionized IGM; the red curve shows the mean $R_\mathrm{p,eq}$ in the case of patchy ionization.  The other curves show the scaling motivated by previous models in the literature.  Patchy ionization has an influence on this scaling, such that the proximity zone sizes for bright quasars are relatively smaller compared to the uniformly ionized case.}
  \label{fig:rpvmagmean}
\end{figure}

The electron collisional ionization rate coefficient values are taken from \citet{1997MNRAS.292...27H}. We use Case~A recombination coefficients \citep{1997MNRAS.292...27H}, which take into account the radiative recombination to all energy levels including the ground state \citep{2009MNRAS.395..736B,2016MNRAS.457.3006D}.  Secondary ionizations can decrease the temperature within the front and somewhat increase the ionization front size in an initial mostly neutral medium \citep{2016MNRAS.457.3006D}. But since our medium is initially mostly ionized, the timescale on which photoelectrons lose their energy through collisional ionizations is $t_{\mathrm{loss}} \propto x_{\mathrm{HI}}^{-1}$ $\sim$ a few hundred Myr, secondary electrons do not play a significant role.  We ignore them in our computation.

The gas temperature is given by
\begin{equation}
  \frac{\ud T}{\ud t} = \frac{2}{3}\frac{\mu m_{\text{H}}}{\rho k_{\text{B}}}\left(\mathcal{H} - \Lambda\right) - 2HT - \frac{T}{n}\frac{\ud n}{\ud t}   
  \label{eq:temp}
\end{equation}
where the heating $\mathcal{H}$ is 
\begin{align}
  \mathcal{H} & = \mathcal{H}_{\mathrm{bg}} + \sum_{i}n_{i}\epsilon_{i} \nonumber \\
  & =  \mathcal{H}_{\mathrm{bg}}  + \sum_{i}\frac{1}{\ud V}\int_{\nu_{i}}^{\infty}\ud\nu\left(h\nu-h\nu_{i}\right)\frac{L_{\nu}}{h\nu}\text{e}^{-\tau_{\nu}}P_{i}.
\end{align}
The heating from background sources was set by assuming thermal equilibrium before quasar turn-on \citep{2012ApJ...746..125H}.  Similar to \citet{2007MNRAS.374..493B}, the cooling term $\Lambda$ includes radiative cooling by recombination \citep{1997MNRAS.292...27H}, free-free emission \citep{1992ApJS...78..341C}, inverse Compton scattering \citep{1971phco.book.....P}, collisional excitation \citep{1997MNRAS.292...27H}, collisional ionization \citep{1992ApJS...78..341C},  cooling due to adiabatic expansion of the universe, as well as due to redistribution of heat between different species. 

Equations~(\ref{eq:HI})--(\ref{eq:HeIII}) and Equation~(\ref{eq:temp}) are coupled and have to be solved numerically.  In order to do this, we discretize each line of sight into uniform cells.  The cell size is fixed and is set by our simulation.  Following \citet{2016MNRAS.457.3006D}, the photoionization and heating rate integrals are evaluated by sampling the frequencies into 80 logarithmic bins between $\nu_{\text{HI}}$ and 40$\nu_{\text{HI}}$.  The equations are updated using a modified version of the implicit Euler method \citep{2013MNRAS.436.2188R} using a fixed global time step that uniformly applies to all cells in a sightline. After solving the thermochemistry equations, we compute the \lya\ optical depth $\tau$ along the line of sight assuming a Voigt absorption profile.  We use peculiar velocities from the underlying hydrodynamical simulation in this process.  The transmitted flux is calculated as $F = \exp(-\tau)$. We smooth the obtained flux with a boxcar filter of 20\;\AA\ and use the definition given by \citet{2006AJ....132..117F} to calculate the proximity zone size as the distance at which the smoothed flux drops below 0.1. Appendix~\ref{app:algo} gives details of our algorithm.  We present results of several code tests in Appendix~\ref{app:tests}.

\begin{figure}
  \includegraphics[width=\columnwidth]{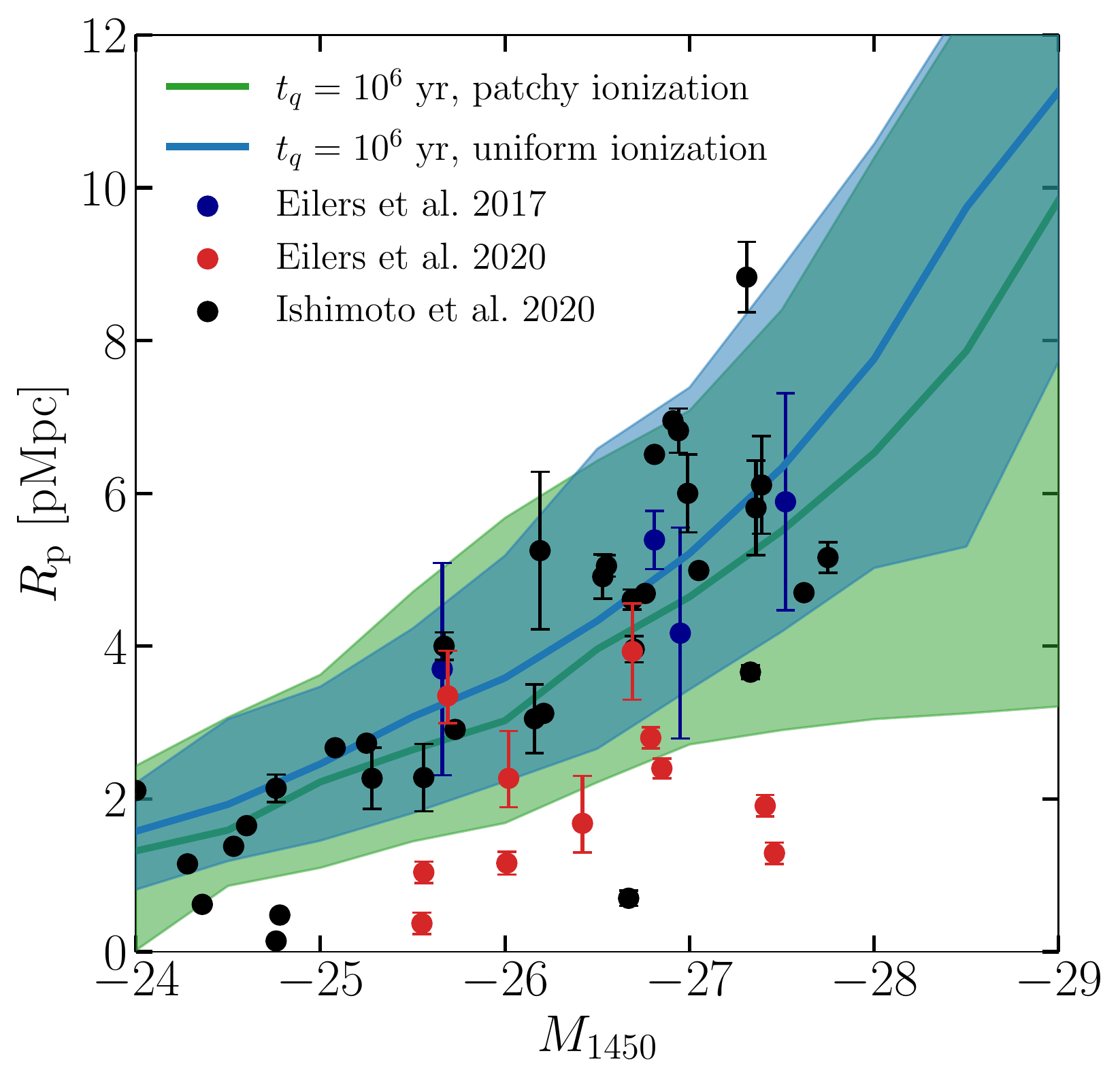}
  \caption{Effect of patchy reionization on proximity zone sizes at $z\sim 6$ in comparison to data. The curves and the shaded regions show the median with 90$\%$ scatter from a sample of 100 sightlines, with the uniform ionization case shown in blue and patchy ionization case shown in green.} 
  \label{fig:withdata}
\end{figure}

Figure~\ref{fig:examplesightline} shows the result of post-processing the sightline at $z = 5.95$ shown in Figure~\ref{fig:3dsnapshots} with the above one-dimensional radiative transfer scheme.  The figure shows the neutral gas density, neutral hydrogen fraction, gas temperature, peculiar velocities, \lya\ optical depth, transmitted flux along the sightline, before the quasar turns on, and 0.1, 1, and 10~Myr after the quasar has turned on.  The quasar is of magnitude $M_{1450}=-26.4$ and has a power law spectral index of $\alpha=-1.7$.  We assume that the quasar is on throughout its lifetime $t_\mathrm{q}$ with constant luminosity (the `lightbulb model').  The quasar is placed in a halo of mass $6.97\times10^{11}\;\msun$.  At this redshift, the sightline is initially almost fully ionized with a few patches of neutral hydrogen that can be seen at $\sim 5$~pMpc.  This results in a temperature of nearly $\sim 10^4$ K everywhere except in the neutral regions.  Given the high cross-section for absorption of the \lya\ line, a neutral fraction of $10^{-4}$ is sufficient to entirely absorb radiation along the sightline before the quasar turns on.  The flux $F$ is therefore uniformly zero at $t_\mathrm{q}=0$.  After the quasar has turned on, the quasar radiation further ionizes the gas as recombinations are negligible ($\alpha~\propto~T^{-0.7}$), resulting in a decrease of neutral hydrogen fraction, while photoionization concomitantly heats up the gas.  As the number of ionizations increase, the recombination rate increases and eventually an equilibrium is reached where the quasar and background radiation balance the recombinations.  This results in a temperature of nearly $\sim 10^4$ K almost everywhere. The proximity zone size $R_\mathrm{p}$ shown at each of the quasar ages $t_\mathrm{q} = 0.1$, 1, and 10~Myr thus increases with lifetime, making it sensitive to lifetime measurements until an equilibrium is reached.

\begin{figure}
  \includegraphics[width=\columnwidth]{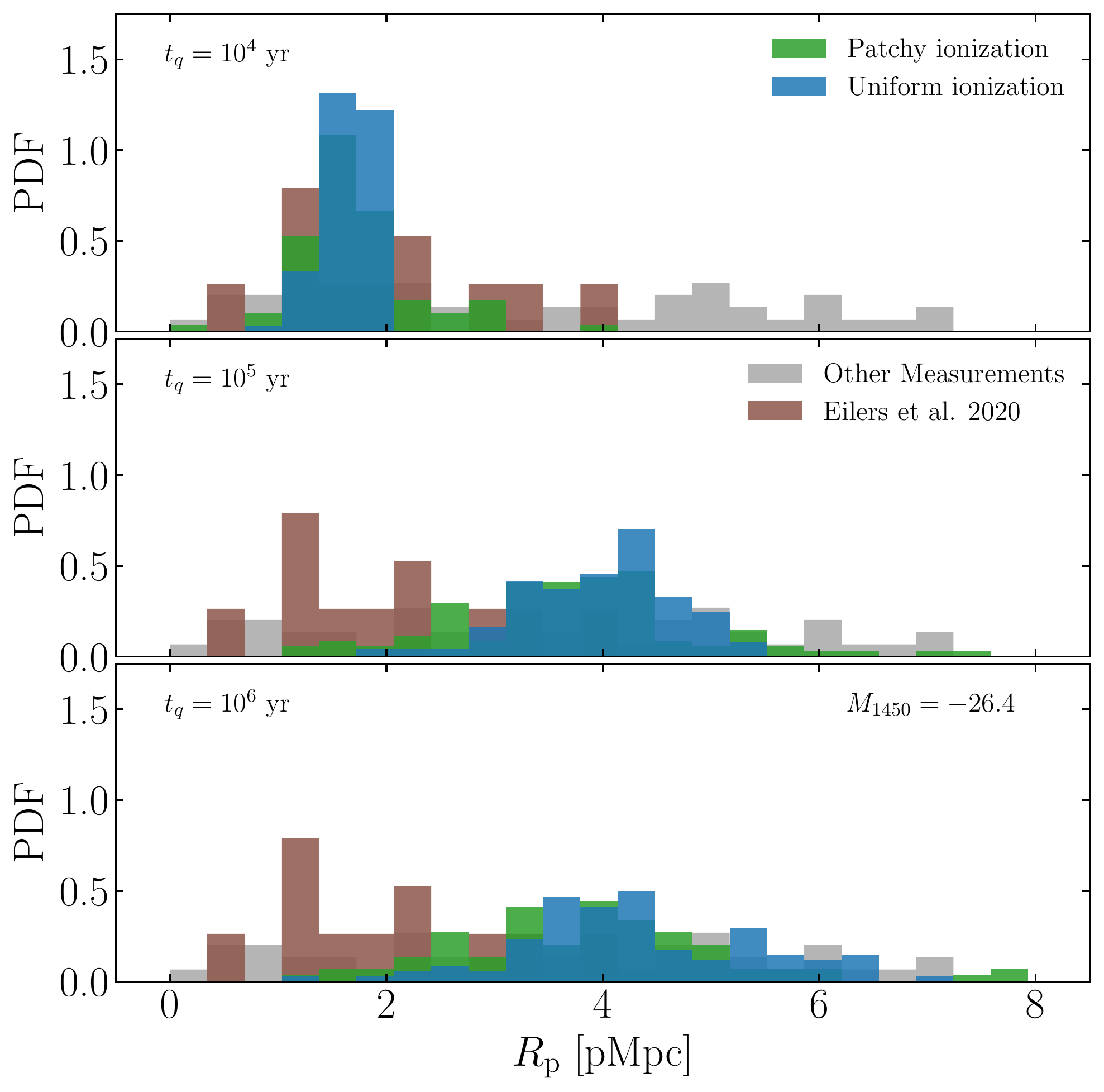}
  \caption{Effect of patchy reionization on proximity zone sizes at $z\sim 6$ in comparison to data. The brown histogram shows results from the targeted survey by \citet{2020ApJ...900...37E}. The grey histogram shows the other data \citep{2017ApJ...840...24E,2020ApJ...903...60I}, which form a homogeneous untargeted sample.\label{fig:withdata2}}
\end{figure}

We test the numerical convergence of our $R_\mathrm{p}$ computation by projecting the density field on grids with successively finer spatial resolution.  We find that the proximity zone size computed at our base resolution of 160--2048 is converged to a relative error of 15\%, with the proximity zone sizes being smaller at higher resolution.

\section{Proximity Zone Sizes}
\label{sec:results}

The computational set-up described above now allows us to discuss the effect of the ionization and cosmological environment of quasars as well as quasar variability on the quasar proximity zones.

\subsection{Effect of reionization topology on proximity zone size}
\label{sec:inicondtns}

Previous studies \citep{2017ApJ...840...24E, 2020MNRAS.493.1330D, 2015MNRAS.454..681K} have assumed that quasar proximity zones grow in an environment that is either completely neutral or already ionized by a background of hydrogen ionizing radiation.  In the latter case, these models set the initial ionization fraction of IGM by invoking a homogeneous UV background.  The resultant ionization distribution therefore follows the cosmological gas density distribution and has values around $x_{\mathrm{HI}}\sim10^{-4}$.  The CROC simulations \citep{2021ApJ...911...60C}  also study proximity zones in the presence of an inhomogeneous ionization background at redshift $\sim$ 6.
  However, they consider the reionization to be mostly complete by redshift  $z \sim 6$. \citet{2007ApJ...670...39L}  consider effects of patchy reionization at $t_{q}= 1$ Myr and conclude that it can lead to longer proximity zone size than in a uniformly ionized medium since quasars are likely to be born inside massive halos that reionize earlier than a typical region. They also point out that the huge sightline-to-sightline scatter in such models might lead to misinterpretations about the ionization state of IGM from  $z\sim 6$ quasar observations. 
Our goal here is to examine if inhomogeneous reionization resulting from late-end reionization models such as ours can result in small proximity zone sizes even for longer quasar lifetimes or lead to longer proximity zone sizes for smaller lifetimes. To study the effects of inhomogeneous reionization on proximity zones, we consider quasars situated in the most massive halos in our simulation with masses between $10^{11}\;\msun$ and $10^{12}\;\msun$.  All quasars are assumed to have $M_{1450}=-26.4$ corresponding to $\dot N=1\times 10^{57}$~photons/s.

\begin{figure}
  \includegraphics[width=\columnwidth]{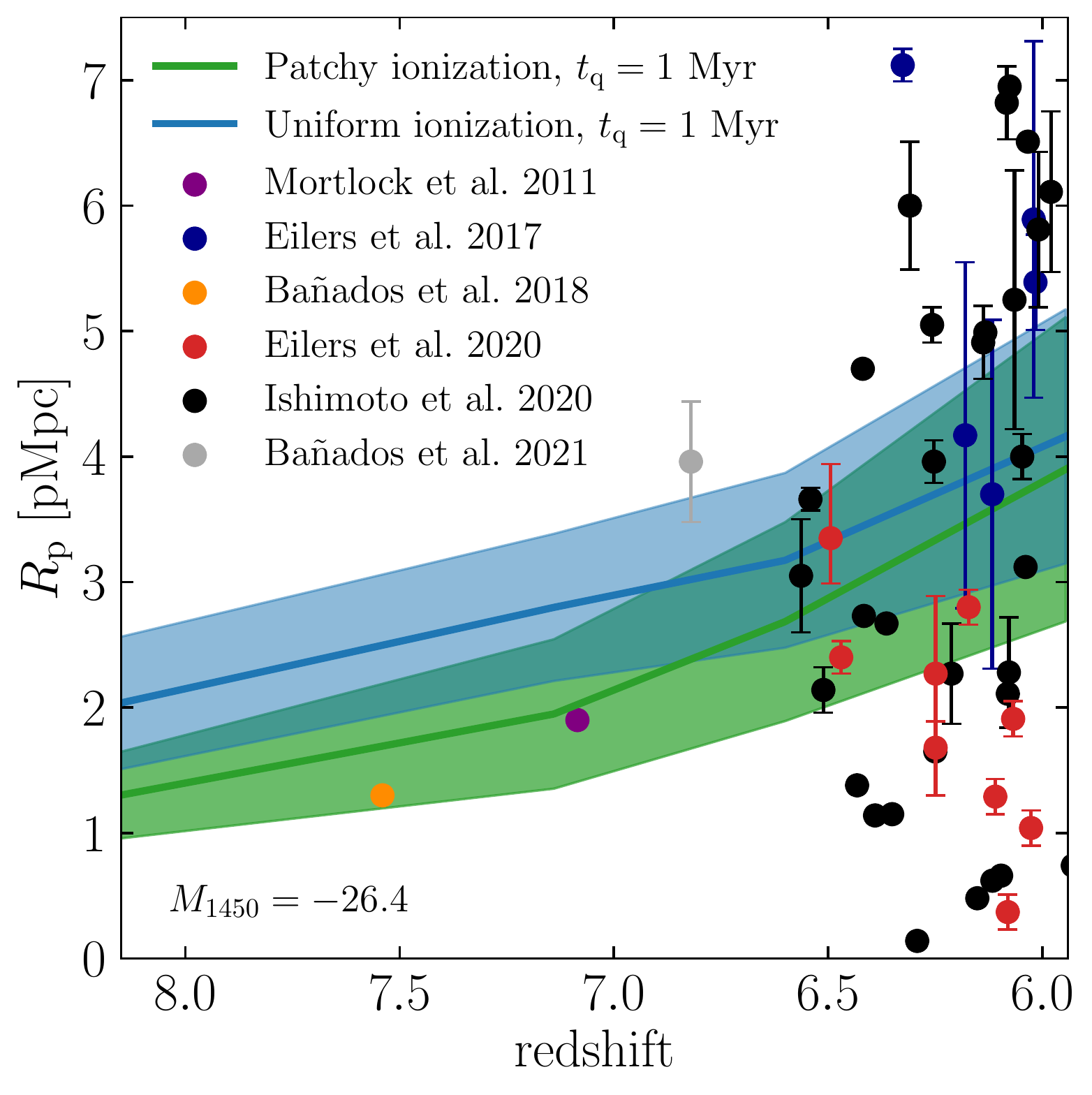}
  \caption{Evolution of the proximity zone size $\rp$ as a function of redshift for quasar magnitude $M_{1450}=-26.4$ and age 1~Myr for the lightbulb model. The curve shows the mean from a sample of 100 sightlines. Shaded region shows the $1\sigma$ (68.26$\%$ equal-tailed credible interval) scatter across sightlines.}
  \label{fig:rpvz}
\end{figure}

Figure~\ref{fig:ini} shows the median evolution of the proximity zone size $R_\mathrm{p}$ for 100 sightlines, for different initial ionization conditions around the quasar, at $z=5.95$ and $z=7.14$.  The figure also shows the 1$\sigma$ spread in the proximity zone sizes (68.26\% equal-tailed credible interval).  Our fiducial computation uses the initial ionization and temperature values for the IGM from our underlying radiative transfer simulation (see Section \ref{sec:simulations}).  We also compare this with a scenario in which the IGM is initially uniformly ionized.  In this case, the ionized hydrogen fraction is assumed to be $x_\mathrm{HI}=10^{-4}$ throughout the box, with a temperature of $T=10^4$~K.  Finally, we consider a case in which the IGM is initially completely neutral, so that initially $x_\mathrm{HI}=1$ and $T=10$~K throughout the box.  The overall evolution shows well-known behavior: $R_\mathrm{p}$ initially increases due to the small photoionization timescale and then becomes constant as ionization equilibrium is reached as recombinations increase and become equal to ionizations.  This is followed by a slight increase in $R_\mathrm{p}$ at later times, which is because of the larger photoionization timescale of \HeII\ and \HeIII\ that delays their ionization.  The associated increase in temperature leads to a decrease in recombination rate and neutral hydrogen fraction, which results in increased transmitted flux.  We see that the initial ambient ionization conditions have a large effect on the evolution of $\rp$.  When the initial ionization state of the IGM is fully neutral, proximity zones are smaller and grow slowly owing to the damping wing as seen in Figure~\ref{fig:uniform_v_aton_compare_panels}.  However, it is also interesting that patchy ionization conditions also have an effect on the growth of $\rp$.  The difference between the fully ionized and the patchy ionized cases is relatively stronger at $z=7.14$, when reionization is half complete in our model.  In the patchy ionization case, the upper bound of the $\rp$ distribution reaches equilibrium sooner, while the lower bound takes longer than a few Myr to equilibrate.  There is a steeper increase in $R_{p}$ post-equilibrium.  The difference between the patchy and uniform cases is also clearly seen in Figure~\ref{fig:uniform_v_aton_compare_panels}, which shows two example sightlines at redshifts $z=5.95$ (left panel) and $z=7.14$ (right panel).  Green, blue, and orange curves show various quantities from the cases with a patchy initial ionization, uniform initial ionization, and no initial ionization.  It can be clearly seen that patches of intergalactic hydrogen along the sightline impede the growth of the proximity zone.

\begin{figure}
  \includegraphics[width=\columnwidth]{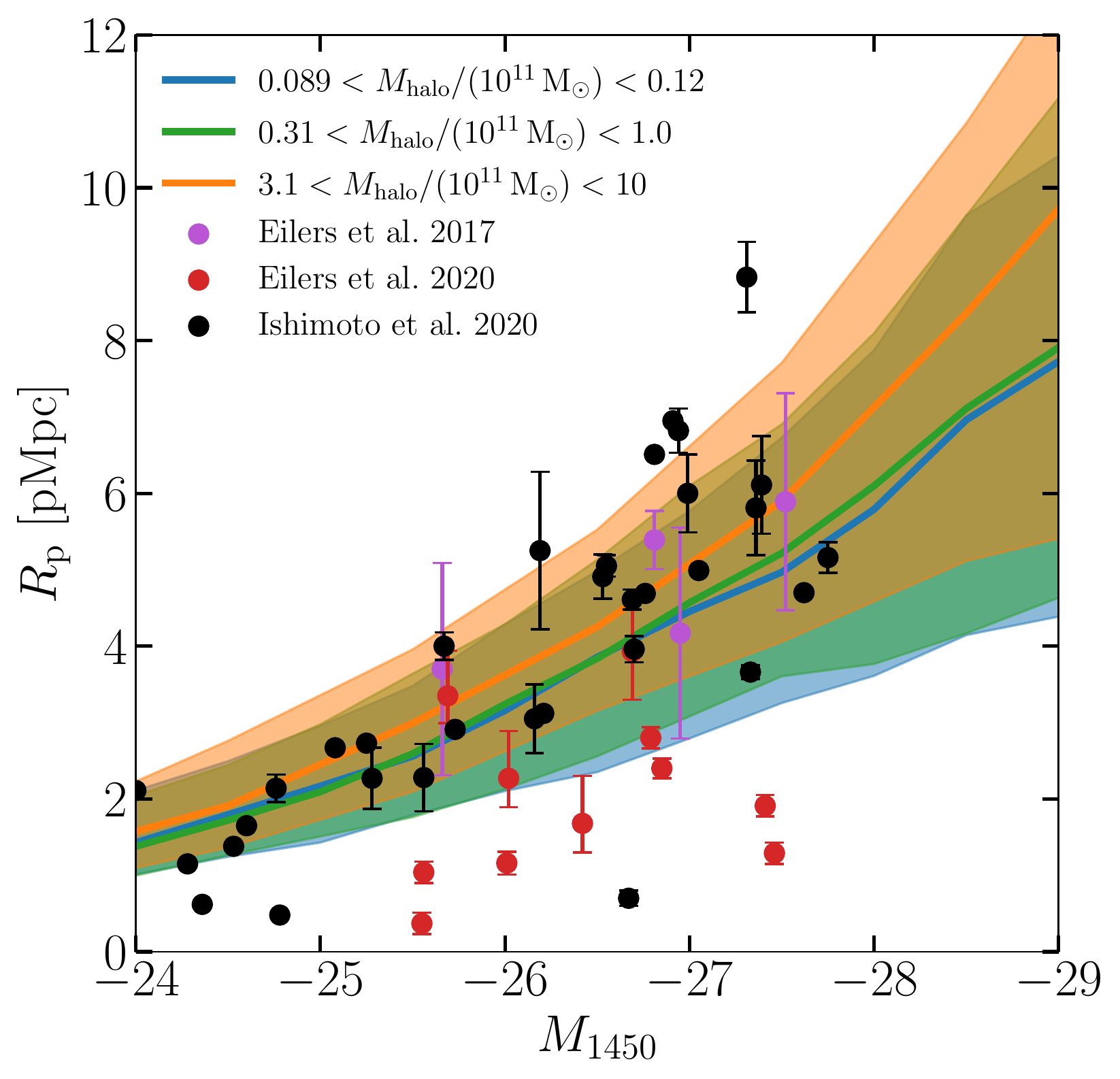}
  \caption{The median proximity zone size $\rp$ and its associated 1$\sigma$ scatter at $z=5.95$ when quasars are placed in one of our three chosen halo mass bins.  The quasar age is fixed at 1~Myr here, and 400 sightlines are used in each case.  The proximity zone size is not strongly affected by the host halo mass, although there is a weak preference for smaller proximity zones around smaller halos due to patchy reionization, as seen in Figure~\ref{fig:hosthalocdf}.}
  \label{fig:rpvmaghalomass}
\end{figure}

It is instructive to investigate the dependence of the equilibrium proximity zone sizes $R_\mathrm{p,eq}$ on quasar magnitude $M_{1450}$ and redshift $z$.  For the case shown in Figure~\ref{fig:ini}, we can safely assume equilibrium at a quasar age of $t_\mathrm{q}=1$~Myr. Once equilibrium is reached, the ionization front radius in a homogenous medium will be equal to the Stromgren radius  
\begin{equation}
  R_\mathrm{ion} =\left(  \frac{3 \dot{N}}{4\pi n_{\text{H}}^2\alpha}\right)^{ 1/3}.
 \end{equation}	
The proximity zone size, $R_\mathrm{p}$, as discussed in the previous section, is defined not by the position of the ionization front but by the point at which the Lyman-$\alpha$ flux crosses an assumed threshold.  But one might still expect $R_\mathrm{p,eq}\propto\dot N^{1/3}$ following the Stromgren argument.  \citet{2007MNRAS.374..493B} derive an analytic estimate for $R_\mathrm{p,eq}$ by considering the flux to be set by the Gunn-Peterson optical depth, 
\begin{multline}
  R_\mathrm{p,eq} =  \frac{3.14}{\Delta_{\text{lim}}} \left(\frac{\dot{N}}{2\times 10^{57}\;\mathrm{s}^{-1}}\right)^{1/2}  \\
  \times\left(\frac{T}{2\times10^4\;\mathrm{K}}\right)^{0.35} \left(\frac{1+z}{7}\right)^{-9/4} \text{pMpc},
  \label{eq:boltrp}
\end{multline}
where $\Delta_{\text{lim}}$ is the gas density corresponding to the flux threshold used to define the proximity zone. 
\citet{2020MNRAS.493.1330D} use the scaling of effective optical depth derived from their simulations instead of Gunn-Peterson optical depth. They derive an analytical value for equilibrium $R_\mathrm{p}$ as
\begin{equation}
	R_\mathrm{p,eq} = r_\mathrm{b}\left[ \left(\frac{\tau_\mathrm{bg}}{\tau_\mathrm{lim}} \right) ^{1/\alpha}-1 \right]^{-1/2},
\end{equation}
where $\tau_{\text{bg}}$  is the effective optical depth in the absence of quasar and $r_{\text{b}}$ is the distance at which quasar radiation equals the background radiation, derived for their simulations as 
\begin{equation}
r_\mathrm{b} = 11.3 \left(\frac{\Gamma_{\text{bg}}}{2.5\times10^{-13} \text{s}^{-1}} \right) ^{-1/2} \left(\frac{\dot{N}}{1.73\times10^{57} \text{s}^{-1}} \right) ^{1/2}  \text{pMpc}.
\end{equation}
Nonetheless, both of the above estimates suggest a steeper $R_\mathrm{p,eq}\propto\dot N^{1/2}$.  The radiative transfer simulations of \citet{2020MNRAS.493.1330D} suggest a slightly modified dependence of $R_\mathrm{p,eq}\propto\dot N^{1/2.2}$.  Figure~\ref{fig:rpvmagmean} compares these estimates to our simulated results by showing the mean evolution of $R_\mathrm{p}$ as a function of magnitude for different initial conditions. We observe that the $R_\mathrm{p,eq}$ is proportional to $\dot N^{1/2.2}$ in good agreement with \citet{2020MNRAS.493.1330D} in the case of an initially uniformly ionized case.  With patchy ionization conditions, the proximity zone sizes are reduced, preferentially for bright quasars, and the dependence of $R_\mathrm{p,eq}$ on $\dot N$ is much shallower, as illustrated by the red curve in Figure~\ref{fig:rpvmagmean}.  The topology of reionization can thus have a considerable effect on proximity zone sizes.  Note that the effect of patchiness seen in Figure~\ref{fig:rpvmagmean} is strongly redshift-dependent, with the scaling moving closer to the $\dot N^{1/3}$ curve at higher redshifts.  This suggests that the scaling of proximity zone sizes invoked in the literature \citep[e.g.,][]{2017ApJ...840...24E} has limited validity.

\begin{figure}
	
  \begin{center}
    \includegraphics[width=\columnwidth]{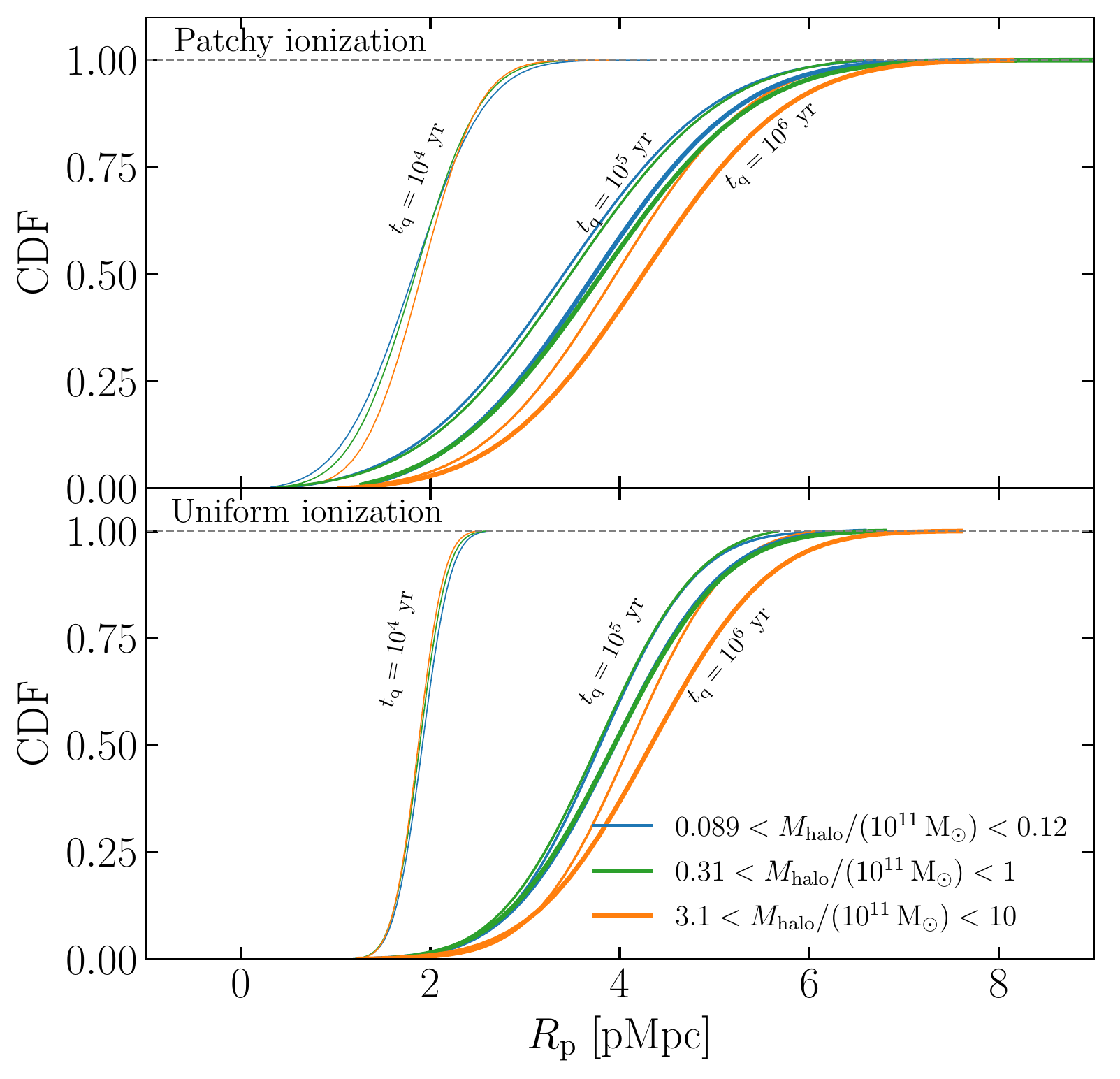}
  \end{center}
  \caption{The cumulative distribution functions of proximity zone sizes at $z=5.95$ for three quasar age values and three choices of the quasar host halo mass.  The quasar magnitude is fixed at $M_{1450}=-26.4$, and 400 sightlines are used in each CDF.  The top panel shows the distributions in the patchy ionization case, while the bottom panel shows the distributions when the IGM is uniformly ionized.  There is an enhancement in the incidence of small proximity zone sizes for small host halo masses.  But this enhancement is significant only in the patchy ionization case, which suggests that it is indirectly caused by small-mass halo environments getting reionized relatively later. \label{fig:hosthalocdf}}
\end{figure}

\begin{figure*}
  \includegraphics[scale=0.45]{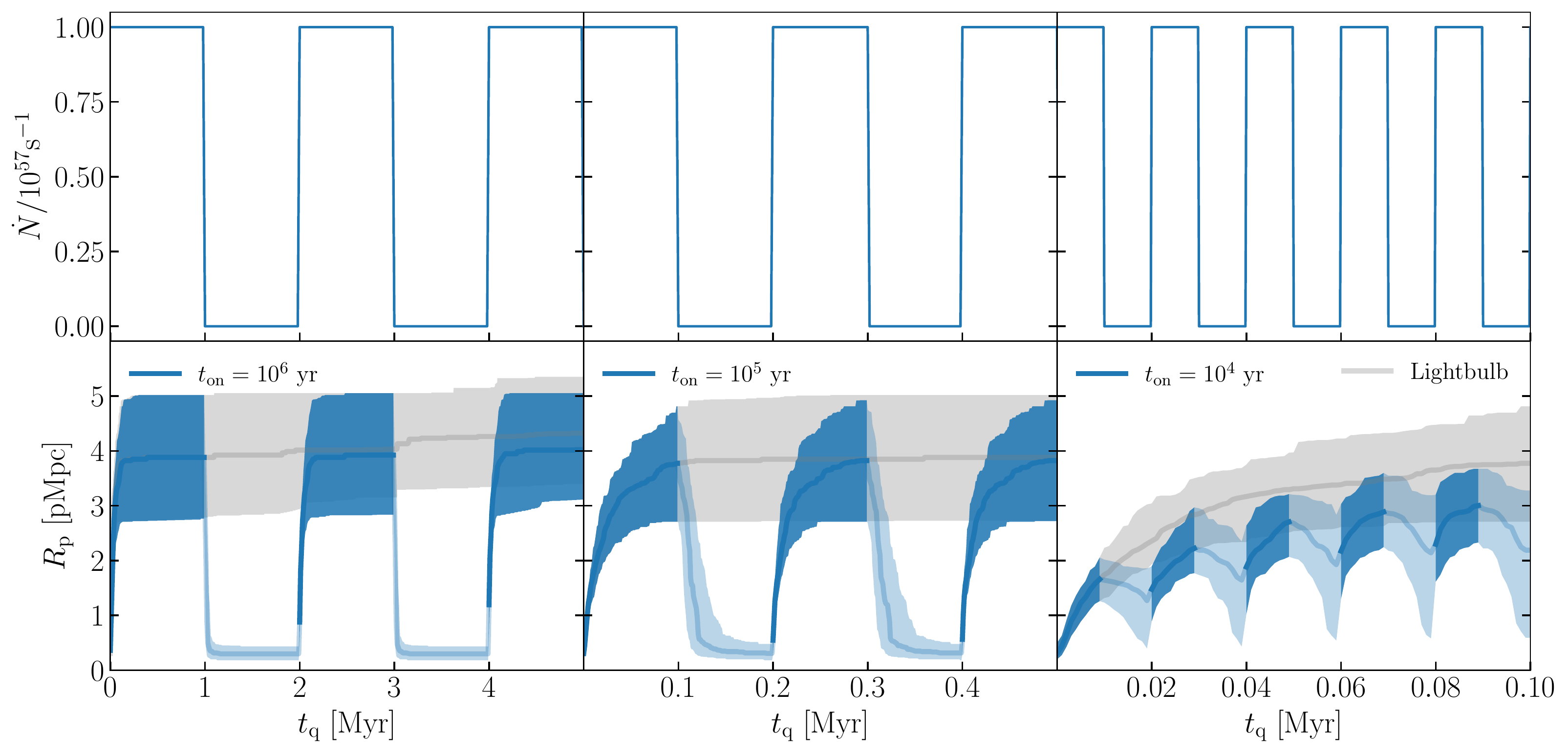}
  \caption{Proximity zone size evolution in periodically varying quasars.  The three columns show quasars that flicker between zero luminosity and $\dot N=10^{57}$~s$^{-1}$ with duty cycle $f_\mathrm{duty}=0.5$ and episodic on time $t_\mathrm{on}=10^6$~yr (left column), $t_\mathrm{on}=10^5$~yr (middle column), and $t_\mathrm{on}=10^4$~yr (right column).  Top panels show the quasar light curves; bottom panels show the corresponding evolution of the proximity zone size.  The blue curves and blue shaded regions in the bottom panels show the median evolution in a sample of 100 sightlines, and the 1$\sigma$ scatter.  Regions shaded in lighter blue in the bottom panels show the proximity zone size evolution during the quasar's off period.  For comparison, the bottom panels also show the proximity zone size evolution for a corresponding lightbulb quasar.}
  \label{fig:episodicrp}
\end{figure*}

In order to examine if this reduction in proximity zone sizes helps reconcile models with data, Figure~\ref{fig:withdata} compares the proximity zone size distributions with measurements.  The data points in this figure span redshifts between $z=5.7$ and $6.5$. \citet{2017ApJ...840...24E} have performed analysis on a homogeneous sample of 34 quasars, both new and archival, to obtain proximity zones. \citet{2020ApJ...900...37E} measured proximity zones by targeting pre-selected quasars that could be potentially young, and performed a multi-wavelength analysis in \citet{2021ApJ...917...38E} to rule out reduction in proximity zone sizes due to proximate DLAs. \citet{2020ApJ...903...60I} have included only quasars with [CII] and Mg II redshifts in their sample, leading to most precise estimates of redshifts and thus proximity zones. They have also updated most measurements from \citet{2017ApJ...840...24E} with the latest redshifts. All of them use the same definition for proximity zone size $R_{\mathrm{p}}$ as discussed in Section \ref{sec:intro}.
 We see that the data are for relatively faint quasars with $M_\mathrm{1450}>-28$ that are affected by reionization topology to a lesser degree as compared to brighter quasars. The median proximity zone size shows only a moderate change between the patchy and ionized cases. However, the enhanced spread in the proximity zone sizes in the patchy model can potentially ease the tension between the models and the data.  
The change in proximity zone sizes because of assuming uniform initial conditions is around 0.29~pMpc at $t = 1$~Myr for a quasar of magnitude $-26.4$, with a maximum change of $0.40$~pMpc considering all quasar ages. Meanwhile, the uncertainties in measured proximity zone sizes can range from $0.14$--$1.43$~pMpc \citep{2017ApJ...840...24E} for redshift 6 quasars.  The uncertainty on $\rp$ due to instrumental noise for fainter quasars is of the same order as redshift errors, while for brighter quasars this error is unknown but expected to be small because of the better signal-to-noise ratio \citep{2020ApJ...903...60I}.  Therefore, both patchy and uniform ionization models are potentially consistent to within the experimental uncertainty at this redshift.  

Figure~\ref{fig:withdata2} shows the reduction in proximity zone sizes in the patchy reionization model for different quasar lifetimes in comparison with data.  The incidence of small proximity zones is greater at longer lifetimes in the patchy ionization model relative to the uniform ionization one. However, in either case, as found previously, the lightbulb model is not sufficient to explain the population of small proximity zones observed.

The reduction in model proximity zone sizes is more significant at high redshifts due to the relatively smaller size of ionized regions around the quasars.  Figure~\ref{fig:rpvz} illustrates this by showing the evolution of the proximity zone size $\rp$ with redshift from our model in comparison with data from \citet{2011Natur.474..616M}, \citet{2017ApJ...840...24E,2020ApJ...900...37E}, \citet{2018Natur.553..473B,2021ApJ...909...80B}, and \citet{2020ApJ...903...60I}.
Analytically, the dependence of $\rp$ on redshift in a uniform density field can be read from Equation (\ref{eq:boltrp}) as $\rp\propto (1+z)^{-2.25}$.  In the \citet{2020MNRAS.493.1330D} model, the redshift dependence comes through $\tau_\mathrm{bg}$ as $(1+z)^{-3.2}$.
In both the uniform and patchy cases, there is reduction in the proximity zone size towards high redshift.  This is partly driven by the density evolution (cf.~Equation~\ref{eq:boltrp}).  However, the evolution is significantly more rapid in the patchy ionization case due to an additional contribution due to the patchiness.  There is also an associated increase in the scatter in the proximity zone sizes.  Early measurements of the proximity zone sizes argued for a rapid evolution between redshifts 5.7 and 6.5 \citep{2010ApJ...714..834C, 2015ApJ...801L..11V} while more recent observations \citep{2017ApJ...849...91M, 2017ApJ...840...24E, 2020ApJ...903...60I} have suggested a shallower trend at the same redshifts.  More data seems to be necessary to measure the average proximity zone size evolution. 

\subsection{Effect of quasar host halo mass on proximity zone size}

The placement of quasars in the cosmological large-scale structure environment is another point on which models have had to make untested assumptions.  This is partly because the host halo masses of $z\sim 6$ quasars are not known, and partly because of the limited dynamic range of simulations.  We now discuss the effect of quasar host halo mass on the size of its proximity zone.  We consider three mass ranges for the host masses $8.9\times 10^9 < M_{\text{halo}}/\mathrm{M}_{\odot}<1.2\times 10^{10}$,  $3.1\times10^{10}<M_\text{halo}/\mathrm{M}_\odot<1.0\times 10^{11}$, $3.1\times 10^{11}<M_\text{halo}/\mathrm{M}_\odot<1.0\times10^{12}$ and examine proximity zone sizes while assuming equal luminosity for all quasars. Overdensities will typically form galaxies sooner, and reionize earlier, so we would expect these regions to be mostly ionized and have larger proximity zones. However, we do not see a significant scaling of the proximity zone size $\rp$ with the host halo mass, as shown in Figure~\ref{fig:rpvmaghalomass}.  This figure shows the median proximity zone size with the associated 1$\sigma$ scatter in a sample of 400 sightlines in each mass bin at $z=5.95$ for a range of quasar luminosities.  The quasar age is fixed at 1~Myr in all cases.  The lack of a dependence of the proximity zone sizes on the halo mass seen in Figure~\ref{fig:rpvmaghalomass} is consistent with previous results \citep{2007ApJ...670...39L, 2015MNRAS.454..681K}. 

The cumulative distribution functions (CDFs) of proximity zone sizes helps understand this.  Figure~\ref{fig:hosthalocdf} shows the CDFs of proximity zone sizes in our three halo mass bins for different quasar lifetimes at $z=5.95$.  CDFs for the patchy ionization case as well as the uniform ionization case are shown.  The quasar magnitude is fixed at $M_{1450}=-26.4$, and 400 random sightlines are used in each case.  We see that the proximity zone sizes in the patchy ionization case are smaller than those in the uniform ionization case, although, as we discussed in the previous section above, this reduction is rather moderate at this redshift.  We also see that in the patchy ionization case, quasars in halos with smaller masses have a greater incidence of small proximity zone sizes.  This is because regions around small-mass halos reionize later \citep{2020MNRAS.491.1736K}, as evidenced by the top panel of Figure~\ref{fig:hosthalocdf}.  Any dependence of the proximity zone sizes on the halo masses is thus indirect and is caused due to the different ionization conditions around halos of different mass.

\begin{figure}
  \includegraphics[width=\columnwidth]{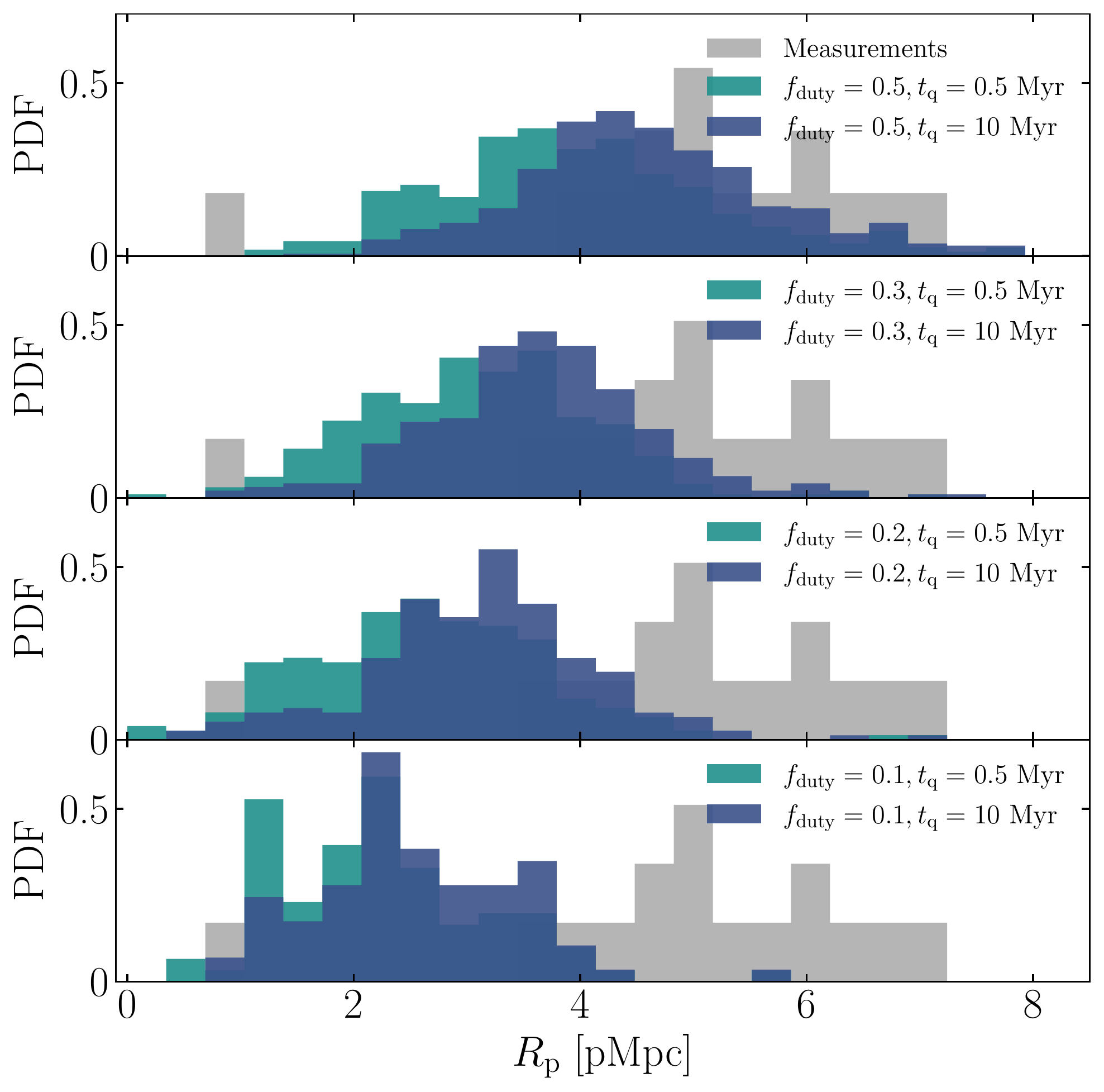}
  \caption{Distribution of the proximity zone size $R_\mathrm{p}$ for a periodically flickering quasar with peak magnitude $M_{1450}=-27$, for various duty cycles and quasar lifetimes.  The episodic on time is held fixed to $t_\mathrm{on}=10^4$~yr.  The proximity zone size is measured only when the quasar is bright.  Grey histograms show the homogeneous sample of measurements by \citet{2017ApJ...840...24E} and \citet{2020ApJ...903...60I}.}
  \label{fig:fdutyhist}
\end{figure}

\subsection{Effect of episodic quasar activity on proximity zone size}

We now investigate the effect of quasar variability on proximity zone sizes.  Observations and simulations suggest that quasars are not constant `lightbulbs'; they flicker on time scales of $\sim 10^{5}$ yr or less \citep{2011ApJ...737...26N, 2013MNRAS.434..606G, 2015MNRAS.451.2517S, 2015MNRAS.453L..46K, 2018MNRAS.474.4740O, 2021ApJ...921...70S}. 
So far, we have been assuming lightbulb quasars in this paper. While realistic light curves will be much more complex and cannot be described by a constant duty cycle, we consider the simpler `blinking lightbulb' scenario, where the quasar periodically turns on for a duration of $t_{\text{on}}$ and off for $t_{\text{off}}$. Once the quasar turns off, the neutral hydrogen fraction will relax to its equilibrium value due to background ionization, on a timescale of $t_{\text{eq}}\sim 1/\Gamma_\mathrm{bg}$.  A quasar can be off either because it is obscured by one of several possible mechanisms, or because the black hole is not accreting.  The subsequent evolution of the proximity zone size is very different from the lightbulb case if this equilibration timescale is shorter than the time for which the quasar is in the off state.  
\citet{2020MNRAS.493.1330D} analytically solved for the behavior of proximity zone sizes in the presence of a flickering quasar with a uniform background radiation. They concluded that the proximity zone sizes are sensitive to episodic lifetime and duty cycle of quasars. Here, we investigate how the distribution of proximity zone sizes changes for varying episodic lifetimes in our patchy reionization simulations.

\begin{figure*}
  \includegraphics[width=1.1\columnwidth]{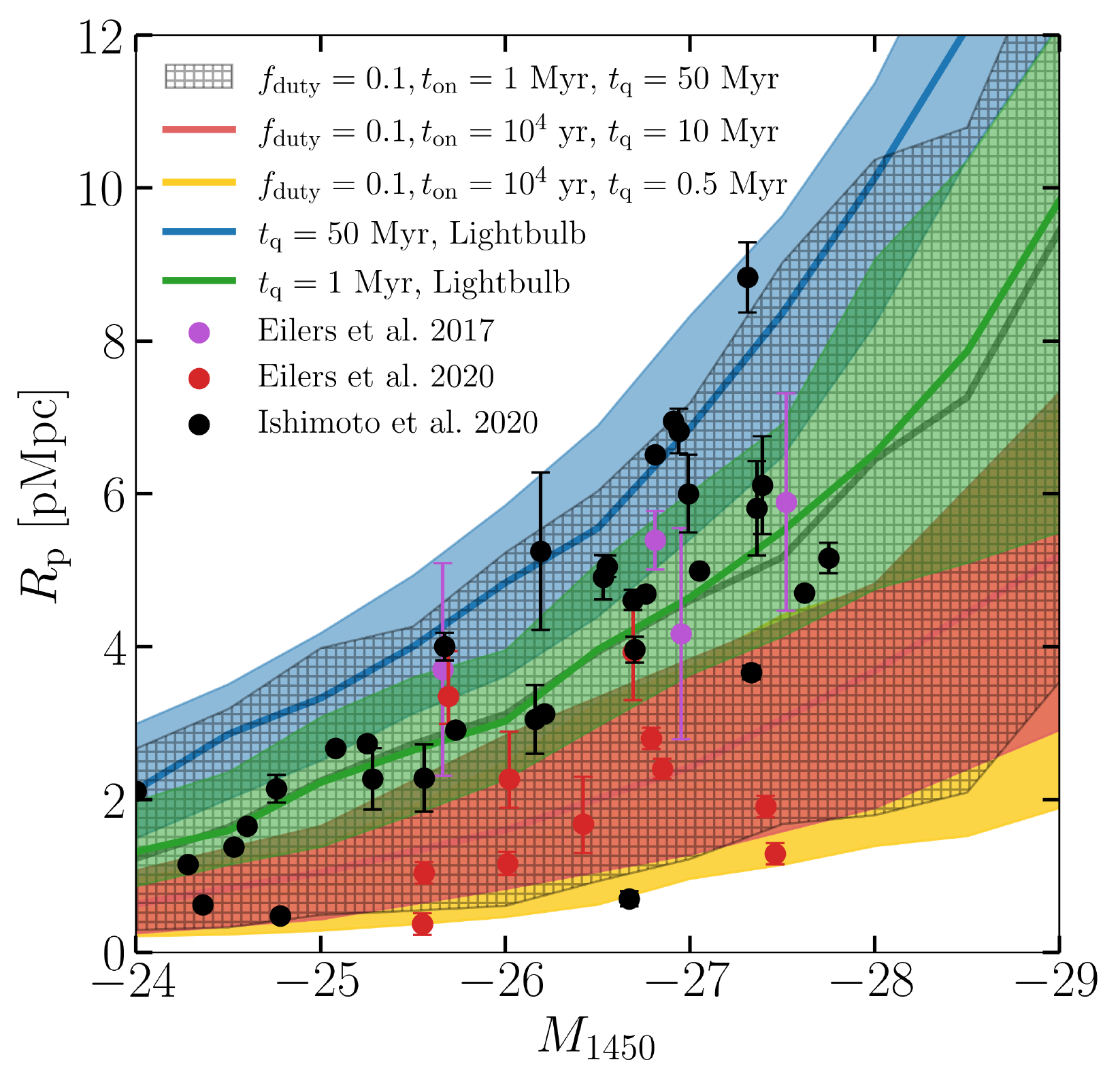}
  \caption{Proximity zone size measurements at $z\sim 6$ by \citet{2017ApJ...840...24E}, \citet{2020ApJ...903...60I}, and \citet{2020ApJ...900...37E}, in comparison with five models.  For each model, we show the median proximity zone size at each quasar magnitude, along with the 1$\sigma$ scatter (68.26\% equal-tailed credible interval).  We randomly sample 100 sightlines from the simulation box, add a random relative temporal offset to each quasar light curve, and use the ionization conditions given by our patchy reionization model. For reasonably large values of $t_{\mathrm{q}}$, large proximity zones can be fit by lightbulb models, while the smaller proximity zones need variable quasars with short episodic on times and small duty cycle.  These variable quasar models continue to fit the small proximity zone sizes well even at longer quasar lifetimes $t_\mathrm{q}$. \label{fig:flicker_with_data}. For a large enough on-time and small enough duty cycle, the hatched region shows our model that is consistent with both small and large proximity zones.}
\end{figure*}

We first consider a simple  model in which the quasar flickers periodically between zero and a fixed luminosity corresponding to $\dot{N}=1 \times 10^{57}\mathrm{s}^{-1}$ ($M_{1450}=-26.4$).  We consider quasars with $t_\mathrm{on}$ between $10^4$ and $10^6$~yr, where the episodic lifetime $t_{\mathrm{on}}$ is the duration of one bright episode in the quasar light curve.  The quasar light curve is assumed to be periodic, so that each cycle is characterized by a bright phase with duration $t_\mathrm{on}$ and an obscured phase with duration $t_\mathrm{off}$.  The duty cycle $f_{\mathrm{duty}}$, defined as the fraction of quasar lifetime that the quasar is on, is then $t_\mathrm{on}/(t_\mathrm{on}+t_\mathrm{off})$.\footnote{Alternative, but related, definitions of the duty cycle are also used in the literature.  For instance, the duty cycle has also been defined as the fraction of the Hubble time for which the quasar is shining \citep{2004ApJ...612..698H}, or the ratio of the number of active and quiescent SMBHs \citep{2012A&A...540A..23S, 2018MNRAS.478.5564B}.}  The top panels of Figure~\ref{fig:episodicrp} show three example light curves describing this scenario.  The quasars are hosted by halos with masses between $10^{11}$ and $10^{12}\msun$.  The initial ionization and thermal state are set by our patchy reionization model.

The lower panels of Figure~\ref{fig:episodicrp} show the evolution of the proximity zone size $\rp$ for such periodic quasars. The shaded region shows the 1$\sigma$ scatter in $\rp$ among 100 sightlines.  The evolution of the proximity zone size during the black hole's bright phase is shown in a bolder color to distinguish it from the evolution during the obscured phase.  The proximity zone size $\rp$ clearly follows the quasar light curve for all three episodic times.

This can be understood as follows. When the quasar is on, the equilibration time for the ionization front, under the assumption of a constant photoionization rate, is \citep{2016ApJ...824..133K}
\begin{equation}
 t_{\mathrm{eq}}^{\mathrm{on}}\gg\frac{1}{\Gamma_{\mathrm{qso}}+\Gamma_{\mathrm{bg}}+n_{\mathrm{e}}\alpha}.
 \label{eq:teq}
\end{equation}
This is the time taken for the gas within the proximity zone to settle in the equilibrium state in presence of the quasar.  

The behaviour of the proximity zone when the quasar is off can be understood as follows.  When the quasar turns off, the number of ionizations is reduced to that only due to background photoionization. If the quasar turns off after an equilibrium is reached, then the number of recombinations are higher than the photoionizations immediately after the quasar turns off. This leads to decrease in the ionization fraction until a new equilibrium is reached between the background photoionizations and recombinations.The timescale to reach this new equilibrium therefore depends on the recombination rate as $t_{\mathrm{eq}}^{\mathrm{off}}\propto 1/\nH\alpha$. On the other hand, if the quasar turns off before an equilibrium is reached between ionizations and recombinations, such that the number of recombinations is still smaller than the background photoionization before quasar turn-off, then post quasar turn-off, the ionized fraction continues to increase, although at a slower rate of $t_{\mathrm{eq}}^{\mathrm{off}} \propto 1/\Gamma_{\mathrm{bg}}$, till it reaches the new equilibrium value. Therefore, the timescale to reach the new equilibrium post quasar turn-off depends on when the quasar turns off once it is turned on.

The timescale for proximity zone to disappear on the other hand depends on the time for the neutral fraction to increase to $\sim 10^{-4}$ once the quasar is turned off, so that the \lya\ absorption is saturated.  We have
\begin{align}
	\frac{\ud\xHI}{\ud t} = -(\Gamma_{\mathrm{bg}}+\Gamma_{\mathrm{qso}})\xHI + (1-\xHI)^2\alpha(T)\nH \\ \approx -(\Gamma_{\mathrm{bg}}+\Gamma_{\mathrm{qso}})\xHI + \alpha\nH.
\end{align}
When the quasar turns off, the equation becomes
\begin{equation}
	\frac{\ud\xHI}{\ud t}  \approx -\Gamma_{\mathrm{bg}}\xHI + \alpha\nH.
\end{equation}
Integrating until a time $t$ after the quasar turns off, the equation becomes
\begin{equation}
	\int_{t_{\mathrm{on}}}^{t_{\mathrm{on}}+t} \frac{\ud\xHI}{-\Gamma_{\mathrm{bg}}\xHI + \alpha\nH} \approx \int_{t_{\mathrm{on}}}^{t_{\mathrm{on}}+t}  \ud t.
\end{equation}
Assuming $\alpha,\nH$ and $\Gamma_{\mathrm{bg}}$ to be constants, the solution can be written as 
\begin{equation}
	\xHI(t) \approx \frac{\alpha\nH}{\Gamma_{\mathrm{bg}}}\left(1-e^{-\Gamma_{\mathrm{bg}}t}\right) + \xHI(t_{\mathrm{on}})e^{-\Gamma_{\mathrm{bg}}t}
\end{equation}
The time $t_\mathrm{vanish}$ at which the proximity zone disappears is such that
\begin{equation}
\xHI(t=t_\mathrm{vanish}) \sim 10^{-4}.  
\end{equation}
 $t_{\mathrm{vanish}}$ can then be computed as 
\begin{equation}
	t_{\mathrm{vanish}} \approx -\frac{1}{\Gamma_{\mathrm{bg}}} \ln{\left(\frac{\Gamma_{\mathrm{bg}}\xHI(t=t_{\mathrm{vanish}})-\alpha\nH}{\Gamma_{\mathrm{bg}}\xHI(t=t_{\mathrm{on}})-\alpha\nH}\right)}
\end{equation}
Assuming $\xHI(t=t_{\mathrm{on}})$ to be $\sim 10^{-8}$ and substituting $\Gamma_{\mathrm{bg}}\sim2.5\times10^{-13}\, \mathrm{s}^{-1}$, $\alpha$ at $T=10^{4}$~K as $4.5\times10^{-13}\,\mathrm{cm}^{3}\mathrm{s}^{-1}$, and an average $\nH\sim10^{-4} \,\mathrm{cm}^{-3}$ at redshift $z \sim 6$, we find
\begin{equation}
  t_{\mathrm{vanish}} \approx 0.1\, \mathrm{Myr}
\end{equation}
For a small $t_{\mathrm{on}}\sim 10^{4}$ yr,  $\xHI(t_{\mathrm{on}})$ can be comparable to $10^{-4}$. Assuming $\xHI(t=t_{\mathrm{on}}) \sim 9\times10^{-5}$, we obtain $t_{\mathrm{vanish}}\approx 0.01$ Myr.  Therefore, $t_{\mathrm{vanish}}$ can have values between $0.01-0.1$ Myr depending on $\xHI(t_{\mathrm{on}})$.  This explains why the proximity zone is destroyed more quickly between cycles than it builds up in the two leftmost panels of Figure~\ref{fig:episodicrp} but not in the rightmost panel.  This allows some proximity zone growth to accumulate over multiple cycles, but as we will see below this growth is not significant for an on time of $10^4$~yr.

If the on-time $t_{\mathrm{on}}$ is greater than the $t_{\mathrm{eq}}^{\mathrm{on}}$ defined in Equation~(\ref{eq:teq}), $\rp$ follows the lightbulb distribution, as seen in the $t_{\mathrm{on}}=10^6$ yr panel in Figure~\ref{fig:episodicrp}.  For $t_\mathrm{on}=10^4$~yr, as in the rightmost panel of Figure~\ref{fig:episodicrp}, the on-time is too short for the proximity zone size to equilibrate to its lightbulb value. Consequently, if in this scenario the duty cycle is small enough, so that $t_\mathrm{off}$ is greater than $t_{\mathrm{vanish}}$, then the proximity zone size remains much smaller than its lightbulb value at all times.  This provides a viable mechanism to explain small proximity zones.

Figure~\ref{fig:fdutyhist} demonstrates this by showing the distribution of proximity zone sizes for a quasar with peak magnitude $M_{1450}=-27$, for various duty cycles and quasar lifetimes.  The proximity zone size distributions are derived only when the black hole is shining and is accessible to the observer.  We show distributions at quasar lifetimes of $t_\mathrm{q}=0.5$~Myr and $t_\mathrm{q}=10$~Myr, to investigate if the small proximity zone sizes vanish at large times.  We randomly sample 100 sightlines from the simulation box, add a random relative temporal offset to each quasar light curve, and use the ionization conditions given by our patchy reionization model.  The episodic on time is held fixed to $t_\mathrm{on}=10^4$~yr, as longer on times will simply take the proximity zone size distribution to the lightbulb value, as we saw in Figure~\ref{fig:episodicrp}.  We see that smaller values of the duty cycle $f_\mathrm{duty}$ yield smaller proximity zone sizes.  This is as expected from our discussion above.  Smaller duty cycles imply longer off times for the quasar, which allows the proximity zone to disappear as the gas in the proximity zone has enough time to equilibrate to the background photoionization rate.  We also see that the proximity zone sizes do not increase significantly even for long quasar lifetimes of $t_\mathrm{q}=10$~Myr.  There is a small increase in $R_\mathrm{p}$ at large $t_\mathrm{q}$ for large values of the duty cycle because these large duty cycles correspond to smaller off times, which prevent complete equilibration.  But there is virtually no change in the distribution of proximity zone sizes for $f_\mathrm{duty}=0.1$ between $t_\mathrm{q}=0.5$~Myr and $t_\mathrm{q}=10$~Myr.  Figure~\ref{fig:fdutyhist} also compares these proximity zone size distributions with the homogeneous sample of measurements by \citet{2017ApJ...840...24E} and \citet{2020ApJ...903...60I}.  We see that models with smaller duty cycles can readily explain even the smallest proximity zones in the data.  It has to be noted that the data histograms do not include the redshift errors on the proximity zone measurements which can go up to 40$\%$. Also, the smallest $\rp$ measurements shown in Figure~\ref{fig:fdutyhist} are at fainter magnitudes compared to our model which explains the slight discrepancy between models and data for $\rp$ less than $\sim$ 1 pMpc. 
 
This picture of variable quasars is put to a more stringent test in Figure~\ref{fig:flicker_with_data}, which aims to model all currently measured proximity zone sizes at $z\sim 6$.  This figure shows measurements by \citet{2017ApJ...840...24E}, \citet{2020ApJ...903...60I}, and \citet{2020ApJ...900...37E}, in comparison with four models.  For each quasar magnitude, we assume a periodic light curve with the given duty cycle and the episodic on time.  As before, we randomly sample 100 sightlines from the simulation box, add a random relative temporal offset to each quasar light curve, and use the ionization conditions given by our patchy reionization model.  (Figure~\ref{fig:flicker_with_data} does not show the proximity zone size measured by \citet{2020ApJ...903...60I} for the quasar J1406--0116.  This quasar shows no Ly$\alpha$ emission line, making it hard to fit a continuum spectrum.  Indeed, \citet{2020ApJ...903...60I} find that the $R_\mathrm{p}$ measurement for this quasar changes significantly, increasing by a factor of $\sim 7$, if the continuum fitting method is changed.)  We see that the large proximity zones are well fit by the lightbulb model at all magnitudes.  These data are therefore also consistent with flickering quasars with long on times.  The smaller proximity zones cannot be fit by lightbulb models with $t_{\mathrm{q}}$ as large as $10^7$ yr.  But these can be fit by models that have a small episodic on time $t_\mathrm{on}=10^4$~yr and small duty cycle $f_\mathrm{duty}=0.1$.  Furthermore, this model continues to describe the small-$R_\mathrm{p}$ data reasonably well even at large quasar lifetimes of $t_\mathrm{q}=10^7$~yr, thus avoiding the need for fine-tuning. For the same model with $f_\mathrm{duty}=0.1$, a larger on-time of $t_\mathrm{on}=10^6$~yr and a lifetime around $t_\mathrm{q}=5\times10^7$~yr is consistent with both small and large proximity zones, with $\rp$ values between 1.2 and 7.1~pMpc at $M_{1450}=-27$, showing that flickering quasars with small duty cycles are not disfavoured by large proximity zones.  These quasars have long unobscured phases, but even longer obscured phases.  Quasar variability thus providing a simple, unified model for proximity zones of all sizes.

\subsection{Consequences for black hole growth}

Although we now see that it is possible to explain the observed  proximity zone sizes via episodic light curves with $t_\mathrm{on}\sim 10^6$~yr and $f_\mathrm{duty}=0.1$, we should now ask whether such a scenario allows formation of black holes with inferred masses by redshifts $z\sim 6$.  This appears to be difficult for the values considered in Figure~\ref{fig:flicker_with_data}.  For example, under the assumption that the accretion rate is proportional to the black hole mass, and assuming a radiative efficiency of $\epsilon=0.1$, a seed of mass $10^3$~M$_\odot$ will require 2~Gyr to grow into a $10^9$~M$_\odot$ black hole with a duty cycle of $f_\mathrm{duty}=1/3$, while accreting at the Eddington rate.  For a duty cycle of $f_\mathrm{duty}=1/10$, equal to what we needed in Figure~\ref{fig:flicker_with_data} for the smallest observed proximity zones, the required time increases to $\sim 7$~Gyr. Not only are these lifetimes greater than those inferred in Figure~\ref{fig:flicker_with_data} by more than an order of magnitude, they are also longer than the age of the Universe at $z\sim 6$ by factors of at least two. As discussed extensively in the literature, growing to masses of $10^9$~M$_\odot$ or more during the optically bright phases requires larger seed masses, larger duty cycles, super-Eddington rates, or a combination thereof \citep{2021ApJ...917...38E}.

One way to alleviate this problem is by having the black hole grow also during obscured phases.  During such phases of obscured growth, the black hole does not shine as a luminous quasar along the observer's sightline.  In this scenario, we can discriminate between a `luminosity duty cycle', $f_\mathrm{duty, lum}$, which quantifies the fraction of the black hole's lifetime for which it shines as an optically bright quasar, and an `accretion duty cycle', $f_\mathrm{duty, acc}$, which is the fraction of the black hole's lifetime for which it accretes and grows.  If $f_\mathrm{duty, lum} < f_\mathrm{duty, acc}$, the black hole undergoes obscured growth, whereas if the two duty cycles are equal, the black hole only grows while it is in the luminous quasar phase.  Using this terminology, the duty cycle $f_\mathrm{duty}$ discussed in the previous section can now be understood as $f_\mathrm{duty, lum}$.

Invoking obscured growth now allows us to solve the black hole growth crisis.  For example, for the cases discussed above, in which we assumed accretion on to the black hole to be proportional to the black hole's mass, a radiative efficiency of $\epsilon=0.1$, a luminosity duty cycle of  $f_\mathrm{duty,lum}=1/3$ or $1/10$, we now find that an accretion duty cycle of $f_\mathrm{duty,acc}=0.7$ readily allows a seed mass of $10^3$~M$_\odot$ at redshift $z=15$ to grow into a $10^9$~M$_\odot$ by $z\sim 6$ within the Hubble time ($\sim 670$ Myr) while accreting at a moderately super-Eddington rate of $\sim$ 1.5.  Pushing the accretion duty cycle closer to unity can even remove the requirement of super-Eddington accretion.

The combination of measurements of quasar proximity zone sizes and the black hole masses thus seem to necessitate obscured growth of SMBHs at high redshifts.  \citet{2019ApJ...884L..19D} and \citet{2021MNRAS.505.5084W} have also argued for obscured black hole growth from their interpretations of proximity zones in hydrogen and helium \lya\ forest spectra.  
For a given observing sightline, obscured black hole growth can occur due to (\textit{a}\/) orientation effects due to dusty torus close to the AGN, so that the black hole continues to accrete and shine, but not along the given sightline \citep{1993ARA&A..31..473A}, or (\textit{b}\/) small-scale physics near the black hole, such as photon trapping, in which photons are unable to escape because efficient accretion of optically thick material impedes photon diffusion \citep{1979MNRAS.187..237B}, or (\textit{c}\/) obscuration by dust, accumulated due to supernovae on the scale of the galaxy \citep{2013Natur.496..329R,2013ApJ...773...44W,2017ApJ...846....3Y,2020ApJ...889...84K}. 
Obscuration is  also likely to occur at a range of different radii at different times due to different mechanisms over the growth history of SMBHs \citep{2015ApJ...802...89B}.  For instance, by cross-correlating the brightest UV-selected AGN from the GOODS sample at redshifts $z\sim 1$--$3$ with X-ray measurements from Chandra, \citet{2017FrASS...4...67D} inferred an obscured AGN fraction of about $0.67$. Recently, \citet{2022arXiv220600018E} observed a heavily obscured quasar at redshift 6.83 over a relatively small COSMOS field, suggesting such quasars might not be too rare at high redshifts. More generally, the obscured AGN fraction is suggested to vary widely between 0.1 and 1 with luminosity \citep{2008ApJ...679..140T, 2015ApJ...802...89B}.  The photon trapping picture has been supported by later analytical and numerical work \citep[e.g.,][]{2000ApJ...539..809Q, 2003ApJ...592.1042I, 2004MNRAS.349...68B, 2015PASJ...67...60T} although some models suggest a reduced efficiency of photon trapping with an associated emission of radiation from polar regions of the accreting black hole \citep{2014ApJ...796..106J}. 

The scenario of photon trapping discussed above is usually associated with super-Eddington accretion and therefore low radiative efficiency. Low radiative efficiency is an alternative to the obscuration scenario, but the required radiative efficiency for a $10^3$~M$_\odot$ seed with $f_\mathrm{Edd}=1$ and duty cycle $0.1$ to grow into a $10^9$~M$_\odot$ black hole within 1~Gyr would be 0.015.  This is consistent with the low radiative efficiencies suggested by \citet{2019ApJ...884L..19D}, but this radiative efficiency is a factor of more than five smaller than theoretical predictions for standard accretion models \citep{1983bhwd.book.....S} and observational measurements \citep{2004MNRAS.354.1020S, 2011ApJ...728...98D,2017ApJ...836L...1T}.

\section{Conclusions}
\label{sec:conclusions}

We have studied the effect of the large-scale ionization environment and quasar variability on high-redshift quasar proximity zones by means of high-dynamic-range radiative transfer simulations of reionization that are calibrated to \lya\ forest measurements.

A key finding of this work is that the topology of residual neutral hydrogen regions in late reionization models can impede the growth of proximity zones.  This leads to a reduction in proximity zone sizes relative to traditional models that mostly assume a fully ionized background.  The reduction is greater for brighter quasars and at higher redshifts. At redshift $\sim 6$, the reduction is greater at lifetimes $ < 10^{7}$ yr. While the particular reionization history considered in this paper does not lead to proximity zone size reduction of a magnitude sufficient to explain the small proximity zones seen in the data, the topology of reionization could potentially be a mechanism to alleviate the tension. For example, in a randomly chosen sample of 100 quasars, we find 12 to have small proximity zones $<2$~pMpc at 1~Myr at redshift $z=5.95$ as opposed to 3 when uniform ionization conditions are assumed around the quasars, for the lightbulb model. The patchiness of reionization has an effect on the slope of the relationship between the proximity zones and quasar magnitude at a fixed quasar age: patchiness makes this relation shallower, so bright quasars have a greater relative decrease in their proximity zone sizes.

In general, the proximity zone size $R_\mathrm{p}$ has a shallow evolution with magnitude. The sightline-to-sightline scatter in $R_\mathrm{p}$ grows with the magnitude.  Proximity zone sizes increase with decrease in redshift and the evolution follows a similar trend as suggested by observations. The sightline-to-sightline scatter increases with decreasing redshift.  Patchy reionization can further enhance the sightline-to-sightline scatter in proximity zone sizes.  Given the quasar luminosity, their proximity zone sizes don't have a strong correlation with host halo mass. However, the fraction of small proximity zones observed within lower mass halos is higher than those in higher mass halos. This is consistent with the idea that these halos are reionized later than the higher mass halos due to having fewer sources.

Recently, \citet{2021MNRAS.508.1853B} have measured an order-of-magnitude reduction in the mean free path of hydrogen ionizing photons at $z=6$ relative to $z=5$.  At $z=6$ the simulation presented in this paper has a larger mean free path compared to the measurement by \citet{2021MNRAS.508.1853B} by about a sigma.  It is plausible that this mismatch stems from small-scale density structure that the simulation fails to resolve.  Resolving this structure may further reduce our predicted proximity zone sizes.  A similar effect that we potentially miss is that of unresolved minihalos with mass $<10^7$~M$_\odot$.  However, such objects are easily destroyed by photoevaporation \citep{2016ApJ...831...86P, 2020ApJ...905..151N} or feedback from star formation \citep{2011MNRAS.417.1480M}.

Although realistic light curves will show a variation of duty cycles during SMBH growth, we assume a simple model with fixed duty cycle and show that quasar variability results in a reduction in proximity zone sizes that can explain the difference between simple lightbulb models and the data, consistent with previous literature. Thanks to the short equilibration timescale of a few tens of thousands of years in the proximity zone, this effect is enhanced for small duty cycles of $f_\mathrm{duty}\lesssim 0.5$ and small episodic lifetimes of $t_\mathrm{on}\lesssim 10^5$~yr.  This not only provides a means to explain the large as well as extremely small proximity zones seen in the data, but also suggests that proximity zone measurements can potentially place constraints on quasar duty cycles and episodic lifetimes.

Quasar variability with short duty cycles aggravate the challenge faced by black hole formation models of forming billion-solar-mass black holes by redshift $z\sim 6$.  But we have seen that this difficulty can be overcome by invoking obscured phase growth of black holes during which the black hole is growing but not shining optically along a given sightline.  The resultant requirements on obscuration can be quantified and turn out to be consistent with observational measurements of obscured quasar fractions and theoretical work on accretion disks.  This not only paves a path towards reconciling observed black hole masses and proximity zone sizes of high-redshift quasars, but also suggests using proximity zone measurements as a probe of obscured black hole growth.  It is important to note that this picture of obscured growth is not limited to quasars with small proximity zones.  While large proximity zones are consistent with lightbulb quasars, and do not pose a problem for black hole growth, quasars with large values of $R_\mathrm{p}$ are not inconsistent with the obscured growth picture.  The proximity zones of these quasars can also be explained by a lightcurve with a small duty cycle, as long as the $t_\mathrm{on}$ is large enough.  Such quasars have long unobscured phases, but even longer obscured phases.  Accretion models that predict obscured growth via processes such as photon trapping will lead to falsifiable predictions for high-redshift quasar proximity zones.  Constraints on these models from proximity zones can also be conceivably combined with measurements of other consequences of these models, such as the presence of \lya\ nebulae \citep[e.g.,][]{2019ApJ...887..196F, 2017ApJ...848...78F, 2018MNRAS.473.3907A, 2022arXiv220311232C}, to gain a fuller picture of how black holes grow to a billion solar masses by redshift six.

\section*{acknowledgments}

We thank James Bolton, Huanqing Chen, Fred Davies, Basudeb Dasgupta, Anna-Christina Eilers, Joe Hennawi, Tom\'a\v{s} \v{S}oltinsk\'y  and Shikhar Asthana for useful discussions, and the anonymous referee for comments that improved the paper.  We also thank Rishi Khatri for computing resources.  GK is partly supported by the Department of Atomic Energy (Government of India) research project with Project Identification Number RTI~4002, and by the Max Planck Society through a Max Planck Partner Group.  This work was supported by grants from the Swiss National Supercomputing Centre (CSCS) under project IDs s949 and s1114.  LK is supported by the European Union’s Horizon 2020 research and innovation programme under the Marie Skłodowska-Curie grant agreement No. 885990.  This work used the Cambridge Service for Data Driven Discovery (CSD3) operated by the University of Cambridge (www.csd3.cam.ac.uk), provided by Dell EMC and Intel using Tier-2 funding from the Engineering and Physical Sciences Research Council (capital grant EP/P020259/1), and DiRAC funding from the Science and Technology Facilities Council (www.dirac.ac.uk).  This work further used the COSMA Data Centric system operated Durham University on behalf of the STFC DiRAC HPC Facility. This equipment was funded by a BIS National E-infrastructure capital grant ST/K00042X/1, DiRAC Operations grant ST/K003267/1 and Durham University. DiRAC is part of the UK's National E-Infrastructure. Support by ERC Advanced Grant 320596 ‘Emergence’  is gratefully acknowledged. For the purpose of open access, the author has applied a Creative Commons Attribution (CC BY) licence to any Author Accepted Manuscript version arising from this submission.

\section*{Data availability}

The data and code underlying this article will be shared on reasonable request to the corresponding author.

\bibliographystyle{mnras}
\bibliography{refs} 

\appendix

\section{Radiative Transfer}
\label{app:algo}

\renewcommand{\thefigure}{B\arabic{figure}}
\setcounter{figure}{0}
\begin{figure}
	\includegraphics[width=\columnwidth]{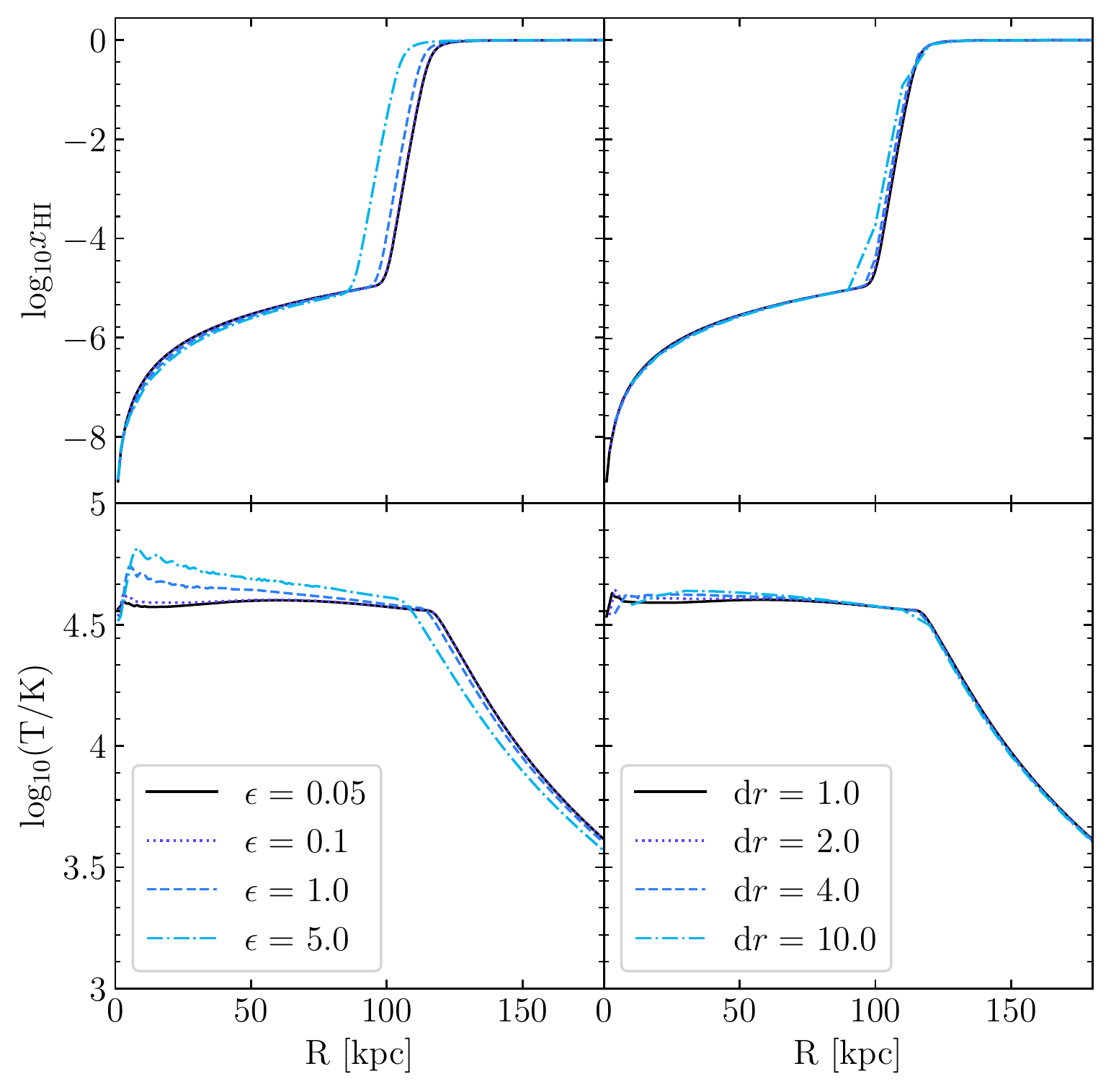}
	\caption{Spatial and Temporal convergence: Top and bottom panels show the abundance of neutral hydrogen and temperature for different cell sizes on the right and different time steps on the left. }
	\label{fig:dtdr_convg}
\end{figure}
\begin{figure}
	\includegraphics[width=\columnwidth]{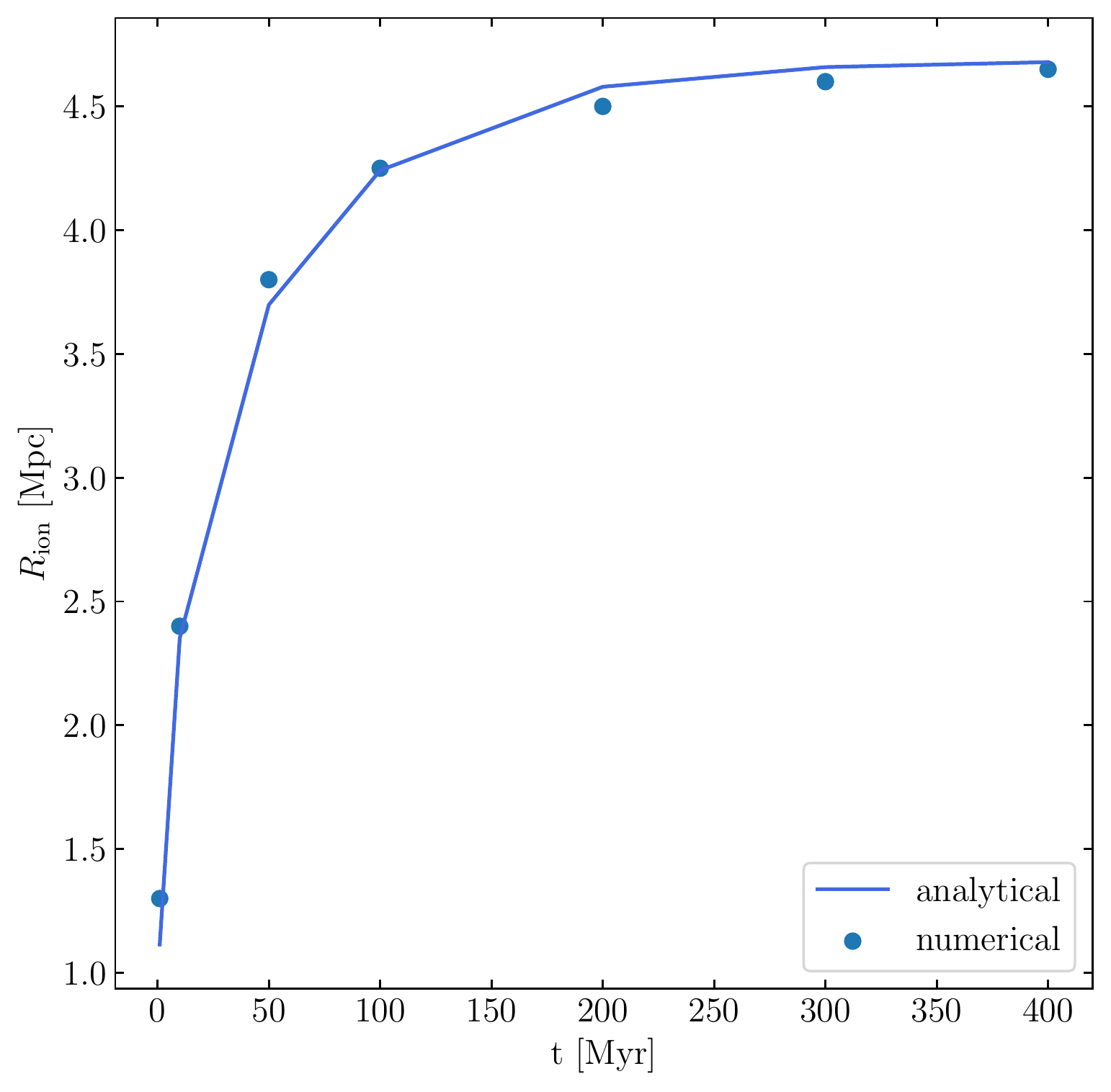} 
	\caption{Comparison between the numerical and analytical Stromgren solutions for ionized radius, for the test parameters described in Section~\ref{sec:stromgren}. Points represent numerical results obtained using our code; blue curve shows the analytical solution.}
	
	\label{fig:bolton_test}
\end{figure}
\begin{figure} 
	\includegraphics[width = \columnwidth]{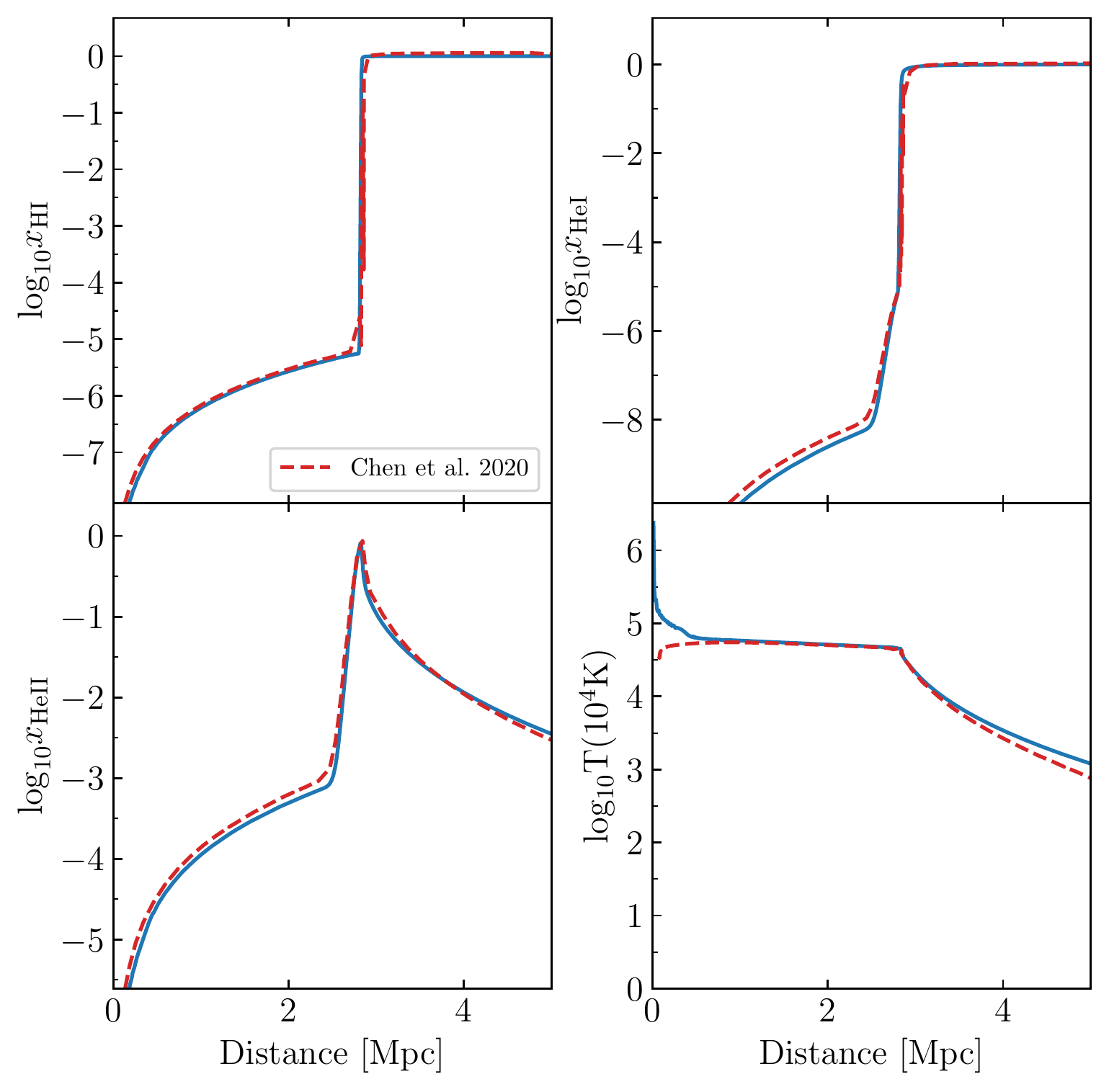}
	\caption{Comparison between results from our code and those from \citet{2021ApJ...911...60C} for the test described in Section~\ref{sec:chen}. The panels show neutral hydrogen fraction, single and doubly ionised helium fractions and temperature. Blue curves are from our code.  Red curves are from \citet{2021ApJ...911...60C}. } 
		\label{fig:chen_test}
\end{figure}
\begin{figure*}
	\includegraphics[scale=0.6]{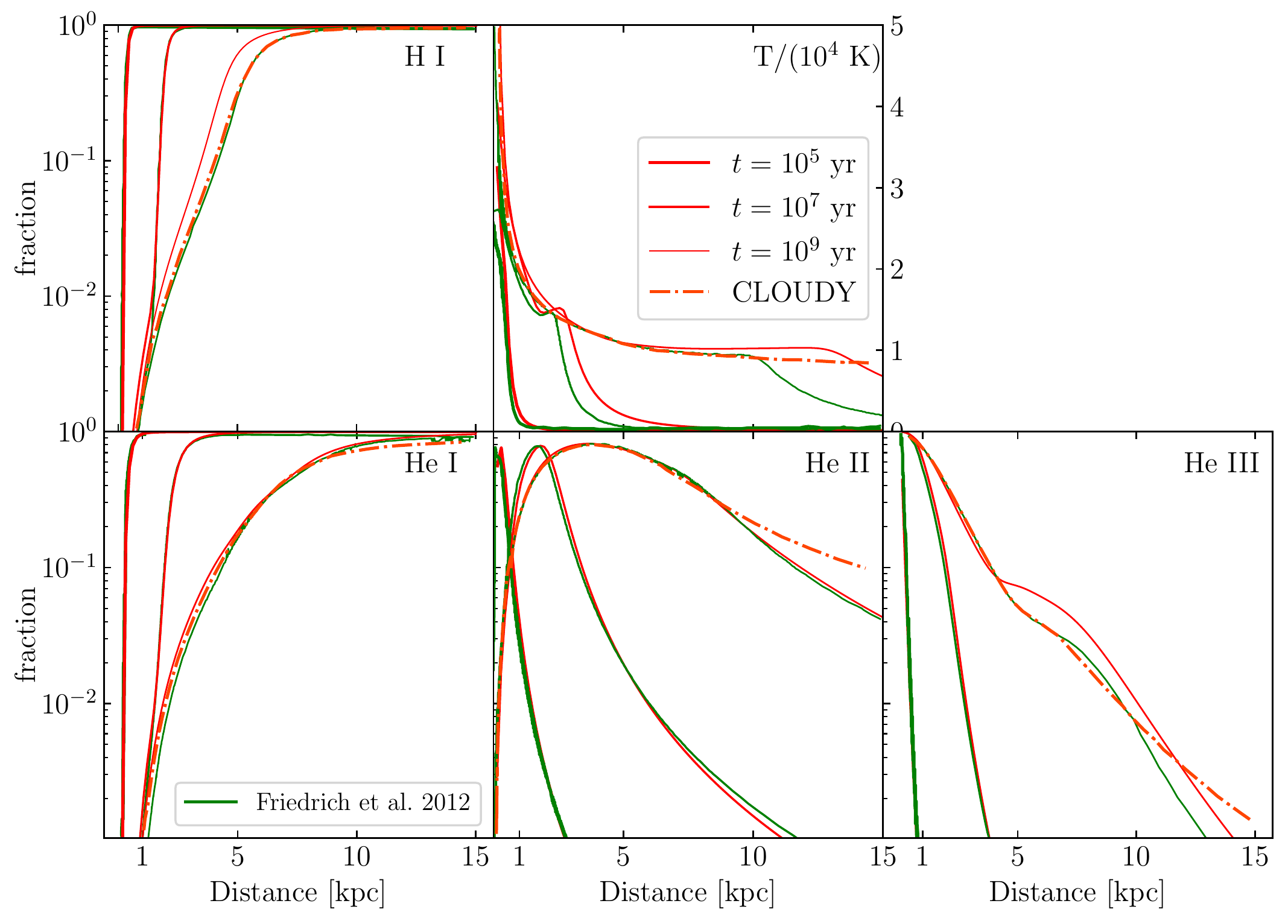}
	\caption{Comparison between results from our code and those from \citet{2012MNRAS.421.2232F} for the test described in Section~\ref{sec:friedrich}. The panels show neutral hydrogen fraction, single, doubly, and triply ionised helium fractions and temperature. Red curves are from our code, green curves are from \citet{2012MNRAS.421.2232F}.  The dot-dashed curves are from CLOUDY.  The thickness of the curves indicates quasar age, as described in the legend. } 
		\label{fig:c2ray_test}
\end{figure*}

We describe our one-dimensional radiative transfer algorithm to solve the thermochemistry equations given by Equations~(\ref{eq:HI})--(\ref{eq:HeIII}) and (\ref{eq:temp}).  The algorithm employs a fixed user-specified global integration time-step.  We exclude the first cell from the computation.

\subsection{Global integration} 

\begin{enumerate}
\item $[\,\mathrm{Initialise\; grid.}\,]$ Set up spatial grid and initialize all quantities of interest (densities, ionization fractions, and temperature) in all the grid cells.
\item $[\,\mathrm{Fix \; global\;timestep.}\,]$ Choose a global time-step $\Delta t$ based on the Courant criterion that depends on cell-crossing time of the ionization front and therefore cell size \citep[cf.][]{2007MNRAS.374..493B}.
\item $[\,\mathrm{Solve\; thermochemistry\; equations\; for \;global\;timestep.}\,]$  Integrate the discretized thermochemistry equations, Equations~(\ref{eq:HI})--(\ref{eq:HeIII}) and (\ref{eq:temp}), following algorithm~\ref{sec:tc}.
\item $[\,\mathrm{Onto\;next\;global\;timestep\; until\; final \;time\;is\;reached.]}$ Repeat step~\textbf{C} with the same global time step until the specified end time is reached.
\end{enumerate}

\subsection{Thermochemistry integration}
\label{sec:tc}

At a global time step, the following algorithm is used to compute the thermochemistry.  This computation is done simultaneously and independently in all cells of the domain \citep[cf.][]{2013MNRAS.436.2188R}.  Radiative quantities such as the photoionization rates, the photoheating rate, and the opacities, are computed when required.  

\begin{enumerate}
\item $[\,\mathrm{Integrate\; Equation\;(\ref{eq:temp}})\,]$ Solve for temperature $T(t+\Delta t)$ using explicit Euler integration with a time step $\ud t = \Delta t$.
\item $[\,\mathrm{Check\; if\; sub\mbox{-}cycling\;is\;required.}\,]$ If $T(t+\Delta t)-T(t)>0.1T(t)$, repeat step~\textbf{A} with smaller time steps $\ud t = \ud t/2$.
\item $[\,\mathrm{ Get \;converged\;} T(t+\Delta t)\mathrm{.\,]}$ Repeat step~\textbf{A} until the condition in step~\textbf{B} is satisfied.
  
\item  $[\,\mathrm{Integrate\; Equation\;(\ref{eq:HI}})\,]$ Solve for $\nHII(t+\Delta t)$ using an implicit Euler scheme, with a time step $\ud t = \Delta t$.  Use $T(t+\Delta t)$ for computing rate coefficients.
\item $[\,\mathrm{Check\; if\; sub\mbox{-}cycling\;is\;required.}\,]$ If $\nHII(t+\Delta t)-\nHII(t)>0.1\nHII(t)$, repeat step~\textbf{D} with smaller time steps $\ud t = \ud t/2$.
\item $[\,\mathrm{ Get \;converged \;}\nHI(t+\Delta t)\mathrm{.\,]}$ Repeat step~\textbf{D} until the condition in step~\textbf{E} is satisfied.
  
\item $[\,\mathrm{Integrate\; Equation\;(\ref{eq:HeI}})\,]$ Solve for $\nHeII(t+\Delta t)$ using an implicit Euler scheme, with a time step $\ud t = \Delta t$.  Use $T(t+\Delta t)$ and $\nHII(t+\Delta t)$ for computing rate coefficients.
\item  $[\,\mathrm{Check\; if\; sub\mbox{-}cycling\;is\;required.}\,]$  If $\nHeII(t+\Delta t)-\nHeII(t)>0.1\nHeII(t)$, repeat step~\textbf{G} with smaller time steps $\ud t= \ud t/2$.
\item  $[\,\mathrm{ Get \;converged \;\nHeII(t+\Delta t)}.\,]$  Repeat step~\textbf{G} until the condition in step~\textbf{H} is satisfied. 
  
\item $[\,\mathrm{Integrate\; Equation\;(\ref{eq:HeIII}})\,]$ Solve for $\nHeIII(t+\Delta t)$ using an implicit Euler scheme, with a time step $\ud t = \Delta t$. Use the updated temperature, $\nHI$ and $\nHeII$ for computing rate coefficients.
\item  $[\,\mathrm{Check\; if\; sub\mbox{-}cycling\;is\;required.}\,]$  If $\nHeIII(t+\Delta t)-\nHeIII(t)>0.1\nHeIII(t)$, repeat step~\textbf{J} with smaller $\ud t = \ud t/2$.
\item $[\,\mathrm{ Get \;converged \;\nHeIII(t+\Delta t)}.\,]$  Repeat step~\textbf{J} until the condition in step~\textbf{K} is satisfied to get converged $\nHeIII$.
  
\item $[\,\mathrm{Compute\; \nel,\nHeI.}\,]$ Compute $\nHeI = \nHe -\nHeII-\nHeIII$ and $\nel=\nHII+\nHeII+2\nHeIII$.
\end{enumerate}

\subsection{Discretization}

Equations~(\ref{eq:HI})--(\ref{eq:HeIII}) are solved in terms of ionization fractions, $x_{\mathrm{HII}}$, $x_{\mathrm{HeII}}$, and $x_{\mathrm{HeIII}}$, instead of the number densities. 
For integrating Equation~\ref{eq:HI}, we follow a semi-implicit numerical scheme as discussed in \citet{2013MNRAS.436.2188R}. The discretized equation looks like
\begin{equation}
 x_{\mathrm{HII}}(t+\Delta t)=   x_{\mathrm{HII}}(t) + \Delta t \frac{C- x_{\mathrm{HII}}(t)(C+D)}{1-J\Delta t}
\end{equation}
where 
\begin{equation}
  J = \frac{\partial C }{\partial x_{\mathrm{HII}}} - (C+D) -x_{\mathrm{HI}}\left(\frac{\partial C }{\partial x_{\mathrm{HII}}}+\frac{\partial D }{\partial x_{\mathrm{HII}}}\right),
\end{equation} 
and $C$ and $D$ are creation and destruction operators that can be read from Equation~(\ref{eq:HI}) after rearranging in terms of $x_{\mathrm{HII}}$ as 
\begin{equation}
  \frac{\ud x_{\mathrm{HII}} }{\ud t} = C - x_{\mathrm{HII}}(C+D).
\end{equation}
A similar semi-implicit Euler scheme is used for integrating Equations~(\ref{eq:HeI}) and (\ref{eq:HeIII}). The discretized equations look as follows
\begin{equation}
  x_{\mathrm{i}}(t+\Delta t) = \frac{ x_{\mathrm{i}}(t)+C\Delta t}{1+D\Delta t}
\end{equation}
where i = HeII, HeIII.  As before, $C$ and $D$ can be read from  Equations~(\ref{eq:HeI}) and (\ref{eq:HeIII}) by rearranging them in the form
\begin{equation}
\frac{\ud x_{\mathrm{i}} }{\ud t} = C - x_{\mathrm{i}}D
\end{equation}
For integrating  Equation~(\ref{eq:temp}), we use an Euler explicit integration scheme. The discretized equation is
\begin{equation}
  T(t+\Delta t) = T(t)  + \Delta t \,L
\end{equation}
where $L$ is the time derivative of temperature evaluated at the previous time step 
\begin{equation}
  L = \frac{2}{3}\frac{\mu m_{\text{H}}}{\rho k_{\text{B}}}\left(\mathcal{H} - \Lambda\right) - 2HT - \frac{T}{n}\frac{\ud n}{\ud t},
\end{equation}
where all the quantities are the same as those in Equation~(\ref{eq:temp}).

\subsection{Source spectrum and photoionization rate}
We assume the source spectrum to be a broken power law \citep{2015MNRAS.449.4204L} given by
\begin{equation}
  L_\nu\propto\begin{cases}
  \nu^{-0.61} & \text{if}~\lambda\ge 912~\text{\AA},\\
  \nu^{-1.70} & \text{if}~\lambda<912~\text{\AA}                
  \end{cases}
\end{equation}
For computing the photoionization rates in Equation~(\ref{eq:Gamma}), we divide the frequencies into 80 bins of equal logarithmic width from $\nu_{\mathrm{HI}}$ to $40\nu_{\mathrm{HI}}$, corresponding to energies between 13.6~eV and 544~eV. Increasing the frequency range or the number of bins does not have any effect on the results. 

\subsection{\lya\ optical depth}

To calculate the \lya\ optical depth, we assume a Voigt profile for the absorption cross-section. The optical depth in a given cell of size $\ud R$ along the line of sight is then calculated as \citep{2018MNRAS.479.2564W}
\begin{equation}
  \tau(i) = \frac{\nu_{\alpha}\sigma_{\alpha}\ud R}{\sqrt{\pi}}\sum_{j<i}\frac{\nHI(j)}{\Delta \nu_{\mathrm{D}}(j)} H(a,x(i,j))
\end{equation}
where $H(a,x)$ is an analytic approximation to the Voigt profile $\phi(\nu)$, related to it as \citep{2006MNRAS.369.2025T}
\begin{equation}
	\phi(\nu) = \nu_{\mathrm{D}}^{-1} \pi^{-1/2} H(a,x),
\end{equation}
and given by 
\begin{multline}
	H(a,x)=\mathrm{e}^{-x^2}- \frac{a}{\sqrt{\pi}x^2} [\mathrm{e}^{-2x^2}(4x^4+7x^2+4+1.5x^{-2}) \\
	-1.5x^{-2}-1 ],
\end{multline}
where $\sigma_{\alpha}$ is the Lyman alpha cross-section, and $\nu_{\alpha}$ the corresponding frequency. The parameter $a$ is
\begin{equation}
	a = \frac{\Lambda_{\alpha}}{4\pi \Delta\nu_{\mathrm{D}}}
\end{equation}	
where
\begin{equation}
\Delta \nu_{\mathrm{D}} \equiv \frac{\nu_{\alpha}}{c}\sqrt{\frac{2k_{\mathrm{B}}T}{m_{\mathrm{H}}}},
\end{equation}
and $\Lambda_{\alpha}$ is the  hydrogen $2p\rightarrow1s$ decay rate.  $x(i,j)$ can be computed given the Hubble velocity $v_{\mathrm{H}}$ and peculiar velocity $v_{\mathrm{pec}}$ within the cell as  
\begin{equation}
	x (i,j) = \sqrt{\frac{m_{\mathrm{H}}}{2k_{\mathrm{B}}T}}[v_{\mathrm{H}}(i)-v_{\mathrm{H}}(j)-v_{\mathrm{pec}}(j)].
\end{equation}

\section{Code Tests}
\label{app:tests}

\subsection{Convergence tests}

To check the spatial and temporal convergence of our algorithm, we run a test from \citet{2007MNRAS.374..493B}. The neutral hydrogen abundance and temperature around a quasar emitting photons at $\dot{N}=5\times10^{53}\text{s}^{-1}$ and having a spectral index of 1.5 is computed after time $t=1$~Myr.  The quasar is assumed to be surrounded by a uniform density medium at redshift $z=6$ with hydrogen and helium in primordial abundance ratio. Figure~\ref{fig:dtdr_convg} shows the results of this test repeated for spatial resolutions $\ud r = 1, 2, 4, 10$~kpc and Courant factors $\epsilon = 0.05, 0.1, 1, 5$. The global time step is fixed by the Courant factor and the spatial resolution as \citep{2007MNRAS.374..493B}
\begin{equation}
  dt  = 3261.6~\text{yr}\;\left(\frac{\epsilon}{0.1}\right)\left( \frac{\ud r}{10 ~\text{kpc}} \right).
  \label{eq:timestep}
\end{equation}
Right column of Figure~\ref{fig:dtdr_convg} shows that the ionization front position and temperature are nearly convergent as cell size becomes less than 2~kpc.  Similarly, the left column of Figure~\ref{fig:dtdr_convg} shows that the ionization front position and temperature are nearly convergent for $\epsilon<0.1$.  Substituting these values in Equation~(\ref{eq:timestep}) gives a global time step of $6.52\times 10^{-4}$~Myr. This time step is four orders of magnitude smaller than the total run time ($\sim 1$~Myr). This can limit the computational time required.  However, in practice, for cosmological conditions, we found that using a global time step that was smaller than the total run time by two orders of magnitude was sufficient to achieve convergence. Similarly, for the flickering quasar, the global time step was chosen to be one-tenth of the episodic time.

\subsection{Stromgren test}
\label{sec:stromgren}

For a monochromatic source in an isothermal and initially neutral uniform hydrogen-only density field, the total number of photons within a shell can be accounted due to ionizations and radiative recombinations as 
\begin{equation}
	\frac{\ud N}{\ud t} = 4\pi R^2  n_{\text{HII}} \frac{\ud R}{\ud t} + \frac{4\pi}{3} R^3n_{\text{HII}}^2\alpha
\end{equation}
Conversely, the rate of change in the ionized volume can be written as 
\begin{equation}
  \frac{\ud R_{\text{ion}}}{\ud t} = \frac{\dot{N}- \frac{4\pi}{3} R^3n_{\text{HII}}^2\alpha}{4\pi R^2  n_{\text{HII}}}
  \label{eq:driondt}
\end{equation}
Equation~(\ref{eq:driondt}) can be solved analytically assuming uniform $n_{\text{HII}}=n_{\text{H}}$ as
\begin{equation}
  R(t) = R_{\text{s}}\left[1- \exp\left(\frac{-t}{t_{\text{rec}}}\right)\right]^{1/3}
  \label{eq:stromgren}
\end{equation} 
where
\begin{equation}
  R_{\text{s}} \equiv \left(\frac{3\dot{\text{N}}}{4\pi \alpha n_{\text{H}}^2}\right)^{1/3},
\end{equation}
and
\begin{equation}
  t_{\text{rec}} = \frac{1}{\alpha n_{\text{H}}},
\end{equation}
and $\alpha$ is the temperature-dependent recombination coefficient. We compare the results from our radiative transfer code with this analytical solution for the test problem 1 from the code comparison project by \citet{2006MNRAS.371.1057I} and find them to be in good agreement. The  ionization front increases in size until the number of recombinations balance out the photoionizations, at which point, the ionization front radius saturates to the Stromgren value $R{\text{s}}$ as shown in Figure~\ref{fig:bolton_test}. 

\subsection{Temperature and Helium evolution}

Analytical tests are not available for the general case in which we solve for the combined hydrogen and helium thermochemistry.  Instead, we test the code by comparing to one-dimensional and three-dimensional results from the literature.  

\subsubsection{Comparison with one-dimensional results by \citet{2021ApJ...911...60C}}
\label{sec:chen}

We solve the thermochemistry and temperature evolution equations in one dimension for a test case from \citet{2021ApJ...911...60C} with the following parameters. The quasar is assumed to emit photons at a rate of $\dot N = 10^{57}\,\text{s}^{-1}$ in a uniform density medium of hydrogen and helium with $X=0.76$ and $Y=0.24$ at a redshift of $z=7$. The source specific luminosity is assumed to be power law with spectral index 1.5. Hydrogen and helium are assumed to be completely neutral initially and the initial temperature before the quasar is turned on is assumed to be 100~K. Figure~\ref{fig:chen_test} shows the neutral hydrogen and ionized helium fractions, as well as temperature at a quasar age of 10~Myr. Our results shown in blue are in good agreement with the results of \citet{2021ApJ...911...60C}, shown in red.

\subsubsection{Comparison with three-dimensional results by \citet{2012MNRAS.421.2232F}}
\label{sec:friedrich}

We compare our results from our one-dimensional radiative transfer code with those obtained using the three-dimensional radiative transfer code C2-RAY by \citet{2012MNRAS.421.2232F}.  The quasar is assumed to emit photons at a rate of $\dot N=5\times10^{48}\text{s}^{-1}$ in a uniform density medium of hydrogen and helium, with $n_{\text{H}} = 10^{-3}\text{cm}^{-3}$ and $n_{\text{He}} = 8.7\times10^{-5}\text{cm}^{-3}$, and an initial temperature of 100~K. The source specific luminosity is assumed to be power law with spectral index 1. Figure~\ref{fig:c2ray_test} shows our results (in orange) for the neutral hydrogen fraction, helium ionization fractions, and gas temperature at quasar ages of $10^5, 10^7,$ and $10^9$~yr against the results from C2-RAY (in green). Also shown are the equilibrium solutions from CLOUDY~\citep{2012MNRAS.421.2232F}. Our results agree very well with the three-dimensional code at $10^5$ and $10^7$ yr. At $10^9$ yr, the helium ionization fractions match very well, but there is a small difference between the hydrogen ionization fronts and temperatures computed by the two codes, potentially due to secondary ionizations that we do not include in our code.

\bsp
\label{lastpage}
\end{document}